\documentclass[11pt,a4paper]{article}
\usepackage[utf8]{inputenc}
\usepackage[round]{natbib} 
\usepackage{epic,eepic,epsfig,amsmath,amssymb,amsthm,enumitem,mathabx}
\usepackage{t1enc,tabularx,subcaption,graphicx}
\usepackage{array}
\usepackage{multirow}
\usepackage{blindtext}

\usepackage[francais,english]{babel}

\usepackage{caption}

\usepackage{color}

\usepackage{bbm}

\usepackage{hyperref}

\def\virgp{\raise 2pt\hbox{,}}

\renewcommand{\geq}{\geqslant}
\renewcommand{\leq}{\leqslant}

\def\R{{\mathbb R}}

\def\virgp{\raise 2pt\hbox{,}}
\def\cdotpv{\raise 2pt\hbox{;}}

\def\1{\mathbbm{1}}

\newtheorem{theorem}{Theorem}[section]

\newtheorem{lemma}[theorem]{Lemma}

\theoremstyle{remark}
\newtheorem{remark}{Remark}[section]

\theoremstyle{definition}
\newtheorem{definition}{Definition}[section]

\theoremstyle{definition}

\theoremstyle{definition}

\newcolumntype{M}[1]{>{\centering}m{#1}}

\begin{document}

\title{The Moroccan Public Procurement Game}

\vskip 0.5cm

\author{Nizar Riane\footnote{Nizar.Riane@gmail.com}}

\maketitle

{\centerline{Faculty of Law, Economic and Social Sciences - Agdal,}
\centerline{Mohammed V University in Rabat,} 
\centerline{Avenue des Nations-Unies, B.P. 721 Rabat, Morocco}

\vskip 1cm

\begin{abstract}
In this paper, we study the public procurement market through the lens of game theory by modeling it as a strategic game with discontinuous and non-quasiconcave payoffs. We first show that the game admits no Nash equilibrium in pure strategies. We then analyze the two-player case and derive two explicit mixed-strategy equilibria for the symmetric game and for the weighted $(p,1-p)$ formulation. Finally, we establish the existence of a symmetric Nash equilibrium in the general $N$-player case by applying an extended version of Dasgupta and Maskin result, which allows us to extend equilibrium existence to the mixed-strategy setting despite payoff discontinuities.
\end{abstract}

\vskip 1cm


\vskip 1cm


\vskip 0.5cm

\noindent \textbf{Acknowledgements}: The author thanks Claire David for her insightful remarks, which greatly improved this work, and Youssef Erraji for engaging in discussions with the author on the functioning of the public procurement market.

\vskip 0.5cm

\section{Introduction}

\hskip 0.5cm In March 2023, the Kingdom of Morocco implemented a significant reform of its public procurement rules for works and service contracts (under the previous system, contracts were awarded to the lowest bidder). According to the decree \citep{DECRET222431}, the \emph{reference price} $P$ is defined as:
\begin{align}
\label{eqn:1}
P &= \frac{E + \bar{x}}{2}
\end{align}

\noindent where $E$ is the estimated cost established by the contracting authority, and $\bar{x}$ denotes the average of all submitted bids.\\

The contract is awarded according to the following rules:

\begin{enumerate}[label=\textit{(\arabic*)}]
\item \label{condition1} Bids must fall within an admissible interval $[A,B]$; otherwise, they are excluded.
\item \label{condition2} The best offer is the one closest from below to the reference price $P$.
\item \label{condition3} If no bid lies below $P$, the winning offer is the closest to $P$ from above.
\item \label{condition4} In the event of a tie, the winner is chosen by lot.
\end{enumerate}

This rule induces a non-cooperative $N$-player game where each bidder chooses a strategy $x_i \in [A,B]$. Because the outcome depends discontinuously in an exotic way on the relative positions of the bids with respect to $P$, the resulting payoff functions are non-smooth, lack convexity, and violate the semicontinuity assumptions required for classical equilibrium existence theorems.\\

From an economic perspective, the new rule transforms the procurement process into a coordination contest similar to a Keynesian beauty contest: bidders attempt to anticipate the collective behavior of others rather than minimizing costs. This feature makes the game both aggregative and discontinuous.\\

Discontinuous games are ubiquitous in economics. In particular, models of oligopolistic competition, such as those of Bertrand and Cournot, exhibit discontinuities in the players' objective functions~(\cite{Reny2020}). Classical results such as Nash's theorem~(\cite{Nash1950}) or Glicksberg's theorem~(\cite{Glicksberg1952}) are of limited use in this context, and one must instead appeal to more equilibrium existence results developed for discontinuous games. For a recent comprehensive review of the literature on discontinuous games, one may consult~\cite{Reny2020}.\\

One of the most important contributions to the study of discontinuous games is the paper by Partha Dasgupta and Eric Maskin~(\cite{Dasgupta1986}). The significance of their result lies in showing how an infinite game can be approximated by a sequence of finite games, thereby providing a constructive method for approximating equilibria.\\

Significant progress has been made in the analysis of symmetric discontinuous games. The first major result was established by~\cite{Baye1993}, who proved the existence of equilibria for discontinuous and non-quasiconcave games satisfying the notions of diagonal transfer continuity and diagonal transfer quasiconcavity. \cite{Reny1999} introduced a similar existence results for quasisymmetric, compact, diagonally quasiconcave games possessing the diagonal better-reply security property. A later and more general contribution was provided by~\cite{BichLaraki2012}, who extended this existence result to games possessing the diagonal local better-reply–correspondence property.\\

In cases where pure strategy equilibria fail to exist, the existence of mixed strategy equilibria can still be established under certain conditions in discontinuous games, but proving properties like better-replay security becomes more challenging.\\

To establish these properties for the mixed extension, several authors have developed conditions under which they are inherited from the pure strategy game to its mixed strategy version. A remarkable contribution by \cite{Monteiro2007} proves equilibrium existence via the notion of \emph{uniform payoff security}, and a closely related concept, \emph{uniformly diagonal security}, was later introduced by \cite{Prokopovych2014}. Although these techniques are often simpler to verify, they can still be too demanding for certain discontinuous games; our public procurement game is one such example.\\

A particularly useful result for verifying payoff security in mixed-strategy games is provided by \cite{Allison2014}. The authors establish a powerful theorem for compact games showing that, if the game satisfies the \emph{disjoint payoff matching} condition, then it is payoff secure.\\

Despite their generality and theoretical depth, it is important to note that these results remain essentially non-constructive, as they do not yield explicit procedures for computing Nash equilibrium strategies.\\

A particularly insightful and constructive approach to proving the existence of equilibria and deriving explicit equilibrium strategies was introduced by Melvin Dresher in~\cite{Dresher1961} and employed for example in~\cite{Hilhorst2018} to solve the $N$-player dual game. This method relies on a system of functional equations that leads to an explicit characterization of the optimal strategy, provided that a suitable form of independence is satisfied in the payoff functions.\\

By situating our problem in context, the Moroccan public procurement game reveals a distinctive structural feature: each player's payoff is a discontinuous non-quasiconcave function that depends not only on their individual strategy but also on an aggregate measure—specifically, the average—of all players' strategies, making the game a particular instance of an aggregative game.\\

According to the definition in \cite{Jensen2018}, a non-cooperative game $\{(S_i, g_i)_{i=1}^N \}$, where each strategy set $S_i \subset \R$, is called \emph{aggregative} if there exists a continuous, additively separable function
\begin{align*}
\phi : \prod_{i=1}^N S_i \to \R
\end{align*}

\noindent (called the \emph{aggregator}) and functions
\begin{align*}
\Phi_i : S_i \times \R \to \R
\end{align*}

\noindent (called the \emph{reduced payoff functions}) such that, for each player $i = 1, \dots, N$ and for all strategy profiles $x = (x_i, x_{-i}) \in \prod_{i=1}^N S_i$,
\begin{align*}
g_i(x_i, x_{-i}) = \Phi_i\big(x_i, \phi(x)\big)
\end{align*}

The distinctive feature of our problem lies in the structure of the payoff function. Although the game is symmetric (see \cite{AlosFerrer2005} for a detailed discussion on symmetric aggregative games), the dependence on the strategy profile is not limited to the individual strategy $x_i$ and an aggregator; instead, it involves the full strategy profile $x$. More precisely, the reduced payoff function is instead of the form
\begin{align*}
\Phi_i : \prod_{i=1}^N S_i \times \R \to \R,
\end{align*}

\noindent so that the payoff for each player $i$ can be written as
\begin{align*}
g_i(x_i, x_{-i}) = \Phi_i\big(x, \phi(x)\big).
\end{align*}

We first examined the public procurement reform in \cite{Riane2026} by modeling it as a Keynesian beauty contest, within the framework of cognitive theory and the common knowledge hypothesis. In that work, we statistically analyzed bidders' behavior using real data from the public procurement market.\\

In this paper, we investigate the theoretical aspect of the problem by formulating it as a strategic game. We study the existence of solutions in both its pure and mixed forms, and we aim to derive these solutions explicitly using constructive methods.\\

The remainder of the paper is organized as follows:\\

In Section~\ref{The Moroccan public procurement game in pure strategies}, we analyze the game and show that it admits no Nash equilibrium in pure strategies.\\

Section~\ref{The two players version} focuses on the two-player case, where we derive two distinct Nash equilibria together with their corresponding mixed strategies.\\

In Section~\ref{The two players $(p,1-p)$ version}, we extend the analysis to a non-symmetric two-player game, allowing for weighted averages.\\

Finally, in Section~\ref{The $N$-players version}, we study the general symmetric $N$-player game and establish the existence of a symmetric Nash equilibrium.\\

\vskip 0.5cm

\section{The Moroccan Public Procurement Game in Pure Strategies\label{The Moroccan public procurement game in pure strategies}}

\hskip 0.5cm It is not difficult to show that the game, as presented in the introduction, does not admit a Nash equilibrium in pure strategies under the rules~\ref{condition1}–\ref{condition4}. Suppose, for contradiction, that a pure-strategy Nash equilibrium exists, denoted by \mbox{$X = (x_1, \ldots, x_N)$}. If at least one player $i$ is not a winner, that player has an incentive to deviate to win, exploiting the sensitivity of the mean to individual bids. More precisely, by playing
\begin{align}
\label{eqn:2}
x^{\star} &= \frac{\sum_{j \neq i} x_j + N E}{2N - 1}
\end{align}

\noindent the player $i$ secure the winning position. In the case of a tie, he can adjust his position to be the unique winner by playing $x^{\star}-\varepsilon$ for some $\varepsilon>0$. In particular, in this degenerate case where all the player play the same strategy $E$, the player can reduce his bid by some $\varepsilon > 0$ to become the closest to the reference price from below, thereby winning the contract under rule~\ref{condition3} and avoiding the tie-breaker in rule~\ref{condition4}.\\

Since the game is symmetric, and defining $g_i$ to be the payoff function for player~$i$, one can remark that
\begin{align}
\label{eqn:3}
0 = \sup_{x_i \in [A, B]} \inf_{x^{-i} \in [A, B]^{N-1}} g_i(x_1, \ldots, x_N)
< \inf_{x^{-i} \in [A, B]^{N-1}} \sup_{x_i \in [A, B]} g_i(x_1, \ldots, x_N) = 1,
\end{align}

\vskip 0.5cm

\section{The Two-Player Version\label{The two players version}}

Does the game admit an equilibrium in mixed strategies? To investigate this, let us consider the two-player version of the game. The payoff function of player~1 is given by
\begin{align}
\label{eqn:5}
g(x,y) &= \mathbf{1}_{\{ y < x \le P \}} 
+ \mathbf{1}_{\{ P \le x < y \}} 
+ \mathbf{1}_{\{ x \le P < y \}} 
+ \frac{1}{2} \mathbf{1}_{\{ x = y \}},
\end{align}
where $\mathbf{1}$ denotes the indicator function, $x$ is player~1's bid, and $y$ is player~2's bid. The winning region is illustrated in Figure~\ref{Graph1}.

\vskip 0.5cm

\begin{figure}[!htb]
\begin{center}
\includegraphics[scale=1]{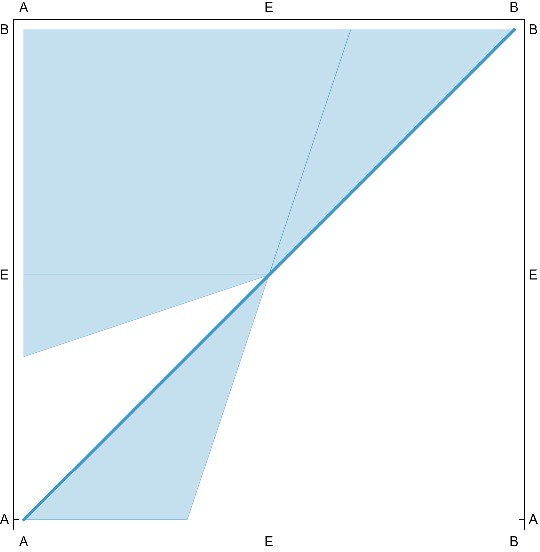}
\end{center}
\captionof{figure}{Payoff function $g$ for two players. The blue region corresponds to a payoff of $1$, the white region to $0$, and the dark blue diagonal to $\frac{1}{2}$.}
\label{Graph1}
\end{figure}

\vskip 0.5cm

Let $\mu$ and $\nu$ be arbitrary probability measures on $[A,B]$, and define $\displaystyle A_i = \frac{A + (3^i - 1)E}{3^i}$. The expected payoff for player~1 can be written in two equivalent forms
\begin{align}
\label{eqn:6}
\begin{split}
G(\mu,\nu) &= \iint g(x,y) \, d\mu(x) d\nu(y)\\
&= \int_A^E \mu\!\Big(\big(y,\tfrac{y+2E}{3}\big]\Big) \, d\nu(y) 
+ \int_{A_1}^E \mu\!\big([A,\,3y - 2E)\big) \, d\nu(y) \\
&\quad + \int_E^B \mu\!\big([A,y)\big) \, d\nu(y) 
- \mu\!\big([A,E)\big)\,\nu(\{E\}) 
+ \frac{1}{2} \sum_{x \in [A,B]} \mu(\{x\})\,\nu(\{x\}).
\end{split}
\end{align}

Alternatively, by changing the order of integration, we have
\begin{align}
\label{eqn:7}
\begin{split}
G(\mu,\nu) &= \iint g(x,y) \, d\mu(x) d\nu(y)\\
&= \int_{A_1}^E \nu\!\big([3x - 2E,\,x)\big) \, d\mu(x) 
+ \int_E^{B_1} \nu\!\big((x,B]\big) \, d\mu(x) \\
&\quad + \int_A^E \nu\!\Big(\big(\tfrac{x + 2E}{3},\,B\big]\Big) \, d\mu(x)
+ \frac{1}{2} \sum_{x \in [A,B]} \mu(\{x\})\,\nu(\{x\}).
\end{split}
\end{align}

In what follows, we present two methods to solve the game, each leading to a distinct mixed-strategy Nash equilibrium. The first method is inspired by the analysis of the counterexample given by~\cite{SionWolfe1958}, while the second applies the approach of~\cite{Dresher1961}.

\vskip 0.5cm

\subsection{The Uniform-Based Strategy\label{The uniform based strategy}}

\hskip 0.5cm We first compute the $\inf \sup G$. Let $\nu$ be an arbitrary probability measure on~$[A,B]$.

\begin{enumerate}
  \item If $\operatorname{supp}(\nu) \subseteq [E,B]$, choose $\mu = \delta_x$ for some $x \in [A,E[$, since the interval~$[E,B]$ is strictly dominated by~$[A,E[$.

  \item If $\operatorname{supp}(\nu) \cap [A,E[ \neq \emptyset$, the mass of~$\nu$ can be partitioned over the intervals $\{[A_i, A_{i+1}[\}_{i \in \mathbb{N}}$. Then:
  \begin{enumerate}
    \item If, for some $i \in \mathbb{N}$,
    \[
    \nu([A_i, A_{i+1}[ \cup [E,B]) \geq \frac{1}{2},
    \]
    choose $\mu = \delta_{A_{i+1}}$.

    \item Otherwise:
    \begin{enumerate}
      \item If $\nu(]A_1, B]) \geq \frac{1}{2}$, choose $\mu = \delta_A$.

      \item If $\nu([A, A_1]) > \frac{1}{2}$ and $\nu([A, A_1[) < \frac{1}{2}$, choose $\mu = \delta_{A_1+\varepsilon}$ for some $\varepsilon > 0$; otherwise, choose $\mu = \delta_{A_1}$.
    \end{enumerate}
  \end{enumerate}
\end{enumerate}

Hence,
\begin{align}
\label{eqn:8}
\sup_{\mu \in \Delta([A,B])} G(\mu,\nu) \geq \frac{1}{2}.
\end{align}

Next, define
\begin{align}
\label{eqn:9}
\nu^{\star}(y) = \frac{1}{2}\Big( U([A, A_1[) + U([A_1, A_2[) \Big)\, \mathcal{L}.
\end{align}

\noindent where $U$ designates the uniform distribution and $\mathcal{L}$ the Lebesgue measure. Then, for all $x \in [A,B]$,
\begin{align}
\label{eqn:10}
G(x, \nu^{\star}) = \frac{1}{2} \, \mathbf{1}_{[A,A_2]} + \frac{A + 26 e - 27 x}{4(E-A)} \mathbf{1}_{]A_2,A_3[} \leq \frac{1}{2}.
\end{align}

This implies that
\begin{align}
\label{eqn:11}
\inf_{\nu \in \Delta([A,B])} \sup_{\mu \in \Delta([A,B])} G(\mu,\nu) = \frac{1}{2}.
\end{align}

One gets
\begin{align}
\label{eqn:10}
G(\mu^{\star},y) = \frac{1}{2} \, \mathbf{1}_{[A,A_2]} + \frac{27 y-5 A-22 E}{4(E-A)} \mathbf{1}_{]A_2,A_3[} + \mathbf{1}_{[A_3,B]} \geq \frac{1}{2}.
\end{align}

Since the game is symmetric, by choosing
\begin{align*}
\mu^{\star}(y) = \frac{1}{2}\Big( U([A, A_1[) + U([A_1, A_2[) \Big)\, \mathcal{L}.
\end{align*}

It follows by a similar argument that
\begin{align*}
\sup_{\mu \in \Delta([A,B])} \inf_{\nu \in \Delta([A,B])} G(\mu,\nu) = \frac{1}{2}.
\end{align*}

Therefore, the game has a value in mixed strategies:
\begin{align}
\label{eqn:12}
\sup_{\mu \in \Delta([A,B])} \inf_{\nu \in \Delta([A,B])} G(\mu,\nu) 
= \frac{1}{2} 
= \inf_{\nu \in \Delta([A,B])} \sup_{\mu \in \Delta([A,B])} G(\mu,\nu).
\end{align}

A Nash equilibrium is thus given by
\begin{align}
\label{eqn:13}
\mu^{\star} = \nu^{\star} = \frac{1}{2}\Big( U([A, A_1[) + U([A_1, A_2[) \Big) \mathcal{L}.
\end{align}

\vskip 0.5cm

\subsection{The Functional Equation Solution\label{The functional equation solution}}

\hskip 0.5cm Suppose that player~2 adopts a strategy with a density~$f$ sufficiently smooth and supported on~$[A,\,\tilde{A}]$. Integrating~$g$ with respect to~$f$ gives
\begin{align}
\label{eqn:14}
\begin{split}
\int_A^B g(x,y)\, f(y)\, dy &= 
\mathbf{1}_{\{ x < E \}} 
\Bigg( \int_{\frac{x+2E}{3}}^{E} f(y)\, dy + \int_{3x-2E}^{x} f(y)\, dy \Bigg)\\
&\quad + \mathbf{1}_{\{ E \le x \}} 
\Bigg( \int_{x}^{3x-2E} f(y)\, dy + \int_{3x-2E}^{B} f(y)\, dy \Bigg).
\end{split}
\end{align}

By the optimality condition and symmetry we have
\begin{align}
\label{eqn:15}
\begin{split}
\int_{\frac{x+2E}{3}}^{E} f(y)\, dy + \int_{A}^{x} f(y)\, dy &= \frac{1}{2}, 
\quad A \le x < A_1,\\
\int_{\frac{x+2E}{3}}^{E} f(y)\, dy + \int_{3x-2E}^{x} f(y)\, dy &= \frac{1}{2}, 
\quad A_1 \le x < E,\\
\int_{x}^{B} f(y)\, dy &= \frac{1}{2}, 
\quad E \le x \le B.
\end{split}
\end{align}

Differentiating these conditions with respect to~$x$ yields the following functional system
\begin{align}
\label{eqn:16}
\begin{split}
-\frac{1}{3} f\!\Big( \tfrac{x+2E}{3} \Big) + f(x) &= 0, 
\quad A \le x < A_1,\\
-\frac{1}{3} f\!\Big( \tfrac{x+2E}{3} \Big) + f(x) - 3\,f(3x-2E) &= 0, 
\quad A_1 \le x < E,\\
-f(x) &= 0, 
\quad E \le x \le B.
\end{split}
\end{align}

The last equation implies that $f$ is zero on~$[E,B]$. Solving the first functional equation on~$[A,\,A_1[$
\begin{align}
\label{eqn:17}
f(x) &= \frac{1}{3} f\!\Big( \tfrac{x+2E}{3} \Big).
\end{align}

Assume $f$ takes the form $f(x) = C_0 (x - E)^s$. Then
\begin{align}
\label{eqn:18}
\frac{1}{3} f\!\Big( \tfrac{x+2E}{3} \Big) 
= C_0\, \frac{1}{3} \Big( \tfrac{x+2E}{3} - E \Big)^s 
= C_0\, \frac{1}{3^{s+1}} (x - E)^s.
\end{align}

Equating gives $s = -1$, so
\begin{align}
\label{eqn:19}
f(x) &= \frac{C_0}{x - E}.
\end{align}

Substituting back into the integral condition for $A \le x < A_1$ and imposing an upper bound~$\tilde{A}$ for the support of~$f$ (to ensure convergence) yields
\begin{align}
\label{eqn:20}
\begin{split}
C_0 &= 
\frac{1}{2} \Bigg( 
\int_{\frac{x+2E}{3}}^{\tilde{A}} \frac{1}{y - E}\, dy 
+ \int_{A}^{x} \frac{1}{y - E}\, dy 
\Bigg)^{-1} \\
&= \frac{1}{2} 
\Big[ \ln(E - \tilde{A}) - \ln(E - x) + \ln(3) + \ln(E - x) - \ln(E - A) \Big]^{-1} \\
&= \frac{1}{2} \ln\!\Big( 3\,\frac{E - \tilde{A}}{E - A} \Big)^{-1}.
\end{split}
\end{align}

Therefore,
\begin{align}
\label{eqn:21}
f(x) &= \frac{1}{2\,\ln\!\big( 3\,\frac{E - \tilde{A}}{E - A} \big)\,(x - E)}.
\end{align}

Imposing the normalization condition
\begin{align}
\label{eqn:22}
\int_A^{\tilde{A}} f(y)\, dy = 1,
\end{align}
we obtain:
\begin{align}
\label{eqn:23}
\ln\!\Big( \frac{E - \tilde{A}}{E - A} \Big) = 2\,\ln\!\Big( 3\,\frac{E - \tilde{A}}{E - A} \Big).
\end{align}

Solving gives:
\begin{align}
\label{eqn:24}
\tilde{A} = A_2 = \frac{A + 8E}{9}.
\end{align}

Knowing the support of~$f$, we can verify the integral condition on~$[A_1,\,E]$, which becomes
\begin{align}
\label{eqn:25}
\int_{3x - 2E}^{x} f(y)\, dy = \frac{1}{2}, \quad A_1 \le x < A_2.
\end{align}

Differentiating again gives
\begin{align}
\label{eqn:26}
f(x) - 3\,f(3x - 2E) = 0,
\end{align}

\noindent which matches the solution of the first functional equation. Thus, we have identified another symmetric Nash equilibrium for the game, distinct from the uniform-based strategy in Section~\ref{The uniform based strategy}. It is given by $\mu = \nu = f\,\mathcal{L}$, where $\mathcal{L}$ denotes the Lebesgue measure and
\begin{align}
\label{eqn:27}
f(x) = \frac{1}{\ln(9)\,(E - x)}\, \mathbf{1}_{[A,A_2]}.
\end{align}

One can check that, for all $y \in [A,B]$,
\begin{align*}
G(x,\nu^{\star}) = \frac{1}{2} \, \mathbf{1}_{[A,A_2]} + \frac{\ln\left(\frac{27 (E-x)}{E-A}\right)}{2\ln(3)} \mathbf{1}_{]A_2,A_3[} \leq \frac{1}{2}.
\end{align*}

\noindent and for all $y \in [A,B]$,
\begin{align*}
G(\mu^{\star},y) = \frac{1}{2} \, \mathbf{1}_{[A,A_2[} + \frac{\ln \left(\frac{E-A}{3 (E-y)}\right)}{\ln (9)} \mathbf{1}_{[A_2,A_3[} + \mathbf{1}_{[A_3,B]} \geq \frac{1}{2}.
\end{align*}

\vskip 0.5cm

\section{The Two-Player \texorpdfstring{$(p,\,1\!-\!p)$}{(p,1-p)} Version\label{The two players $(p,1-p)$ version}}

\hskip 0.5cm In this section, we generalize the symmetric two-player game by introducing an asymmetry in the definition of the reference price. This modification allows us to study how the equilibrium changes when the two bidders do not contribute equally to the determination of the reference value. \\

Let the reference price now depend asymmetrically on the two bids
\begin{align}
\label{eqn:28}
P &= \frac{\, p\,x + (1 - p)\,y + E\,}{2},
\end{align}

\noindent where $p \in [0,1]$ is a weight parameter representing the relative contribution of player 1’s bid to the reference price. The original symmetric case corresponds to $p = \frac{1}{2}$. The payoff function of player~1 becomes
\begin{align}
\label{eqn:29}
g_p(x,y) &= \mathbf{1}_{\{\, y < x \le P \}} 
+ \mathbf{1}_{\{\, P \le x < y \}} 
+ \mathbf{1}_{\{\, x \le P < y \}} 
+ p\,\mathbf{1}_{\{\, x = y \}}.
\end{align}

The discontinuities in $g_p$ are now asymmetric, since player 1's own bid influences the reference price more strongly when $p > \tfrac{1}{2}$, and less strongly when $p < \tfrac{1}{2}$. This asymmetric formulation induces a family of discontinuous zero-sum games parameterized by $p$, each of which has its own equilibrium structure and game value $v(p)$.\\

Before turning to explicit solutions, it is useful to interpret the parameter $p$:

\begin{enumerate}
\item When $p \to 0$: the reference price depends almost entirely on player 2's bid.
Player 1 has negligible influence and thus a weaker strategic position.

\item When $p \to 1$: the situation reverses, and player 1 almost determines $P$ by herself. The game becomes almost trivial, with player 1 effectively playing against her own benchmark.

\item When $p = \tfrac{1}{2}$: the game is perfectly symmetric, corresponding to the benchmark model analyzed in Section \ref{The two players version}.
\end{enumerate}

The parameter $p$ therefore introduces an imbalance of influence over the reference price, leading to a family of games that transition from symmetric to asymmetric regimes.\\

Figure~\ref{Graph2} illustrates the winning region for player~1 in the case $p = 0.1$.

\vskip 0.5cm 

\begin{figure}[!htb]
\begin{center}
\includegraphics[scale=1]{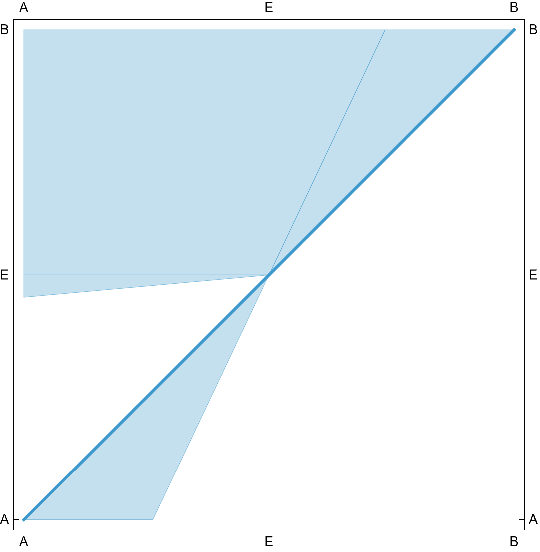}
\end{center}
\captionof{figure}{Payoff function $g_p$ for $p = 0.1$. The blue region corresponds to a payoff of $1$, the white region to $0$, and the dark blue diagonal to~$p$.}
\label{Graph2}
\end{figure}

\vskip 0.5cm

We introduce the maps
\begin{align}
\label{eqn:30}
\begin{split}
f_{1,p}(x) &= \frac{\, p\,x + E\,}{p + 1}, 
\quad 
f_{2,p}(x) = \frac{\, (1 - p)\,x + E\,}{2 - p},\\
h_{1,p}(x) &= \frac{\, (2 - p)\,x - E\,}{1 - p}, 
\quad 
h_{2,p}(x) = \frac{\, (p + 1)\,x - E\,}{p}.
\end{split}
\end{align}

Define the sequences
\begin{align}
\label{eqn:31}
\begin{split}
\hat{A}_i &= f_{1,p}(\hat{A}_{i-1}), 
\quad 
\check{A}_i = f_{2,p}(\check{A}_{i-1}), 
\quad 
\hat{A}_0 = \check{A}_0 = A. \\
\hat{C}_{i} &= h_{1,p}(\hat{A}_i), 
\quad 
\check{C}_{i} = h_{2,p}(\check{A}_{i+1}),\\
\hat{D}_{i+1} &= f_{2,p}(\hat{D}_{i}), 
\quad 
\hat{D}_1 = \frac{\, (p + 1)\,(1 - p)^2\,A + [\, (5 - 3p)\,p - 1\,]\,E\,}{\, p\,(2 - p)^2 },\\
\check{D}_{i+1} &= f_{1,p}(\check{D}_{i}), 
\quad 
\check{D}_1 = \frac{\, 2\,E + p\,(1 - p)\,A\,}{\, (2 - p)\,(p + 1) }.
\end{split}
\end{align}

Note that for any actions $x$ and $y$ by players~1 and~2 respectively, the strict winning region for the player~1 is
\begin{align}
\label{eqn:32}
\begin{split}
w(x) &= [\, h_{1,p}(x),\, x\, [ \; \cup \; ]\, f_{1,p}(x),\, B\, [ ,\\
w(y) &= [\, A, h_{2,p}(y)\, [ \; \cup \; ]\,  y,\, f_{2,p}(y) [ .
\end{split}
\end{align}

One should distinguish five cases:
\begin{enumerate}
\item $p = \tfrac{1}{2}$,

\item $\tfrac{1}{6}\big(5 - \sqrt{13}\big) < p < \tfrac{1}{2}$,

\item $p = \tfrac{1}{6}\big(5 - \sqrt{13}\big)$,

\item $0 < p < \tfrac{1}{6}\big(5 - \sqrt{13}\big)$,

\item $p = 0$.
\end{enumerate}

The threshold value $p = \tfrac{1}{6}(5 - \sqrt{13})$ arises as the solution of the condition:
\begin{align}
\label{eqn:33}
h_{2,p}\!\big(\check{A}_2\big) = A.
\end{align}

Finally, observe that for $p = 0$, the game admits a pure-strategy Nash equilibrium at $x = y = E$, with value~$0$.

\vskip 0.5cm

\subsubsection{The The Intermediate Regime \texorpdfstring{$\frac{1}{6}(5 - \sqrt{13}) < p < \frac{1}{2}$}{(1/6)(5 - sqrt(13)) < p < 1/2}}

\hskip 0.5cm For this range of~$p$, the game is no longer symmetric. Let $\mu$ and $\nu$ be two arbitrary probability measures on~$[A,B]$ and define
\begin{align}
\label{eqn:34}
G_p(\mu,\nu) = \iint_{[A,B]^2} g_p(x,y)\, d\mu(x)\, d\nu(y).
\end{align}

\begin{enumerate}
  \item If $\operatorname{supp}(\nu) \subseteq [E,B]$, player~1 can choose $\mu = \delta_x$ for some $x \in [A,E[$.

  \item If $\operatorname{supp}(\nu) \cap [A,E[ \neq \emptyset$, then by choosing $\mu = \delta_A$, $\mu = \delta_{\hat{C}_1}$, or $\mu = \delta_{\hat{A}_1}$, player~1 guarantees at least~$\frac{1}{3}$. Moreover, if we partition the interval $[A,\,\check{D}_1]$ into the five subintervals 
  \[
  [A,\check{C}_1[,\quad [\check{C}_1,\check{A}_1[,\quad [\check{A}_1,\check{C}_2[,\quad [\check{C}_2,\check{A}_2[,\quad [\check{A}_2,\check{D}_1],
  \]
  then player~1 can secure a win in at most two out of five subintervals by selecting:
  \begin{enumerate}
    \item $\mu = \delta_{\check{D}_1}$: wins on $[\hat{A}_1,\check{A}_2[ \subset [\check{C}_2,\check{A}_2[$ and $[\check{A}_2,\check{D}_1[$.
    \item $\mu = \delta_{\check{C}_1}$: wins on $[A,\check{C}_1[$ and $[\check{A}_2,\check{D}_1[$.
    \item $\mu = \delta_{\check{A}_1}$: wins on $[A,\check{C}_1[$ and $[\check{C}_1,\check{A}_1[$.
    \item $\mu = \delta_{\check{C}_2}$: wins on $[\check{C}_1,\check{A}_1[$ and $[\check{A}_1,\check{C}_2[$.
    \item $\mu = \delta_{\check{A}_2}$: wins on $[\check{A}_1,\check{C}_2[$ and $[\check{C}_2,\check{A}_2[$.
  \end{enumerate}

  Thus, the worst case for player~1 yields a payoff of at least~$\frac{2}{5} - \varepsilon_p$ for some $\varepsilon_p > 0$.
\end{enumerate}

Therefore,
\begin{align}
\label{eqn:35}
\sup_{\mu \in \Delta([A,B])} G_p(\mu,\nu) \ge \frac{2}{5} - \varepsilon_p > \frac{2}{5}.
\end{align}

Next, define
\begin{align}
\label{eqn:36}
\nu^{\star} = 
\Big(
  \check{p}_1\,U([A, \check{C}_1[) 
+ \check{p}_2\,U([\check{C}_1,\check{A}_1[) 
+ \check{p}_1\,U([\check{A}_1, \check{C}_2[) 
+ \check{p}_2\,U([\check{C}_2, \check{A}_2[) 
+ \check{p}_2\,U([\check{A}_2, \check{D}_1[)
\Big) \mathcal{L}.
\end{align}

Then, for all $\mu \in \Delta([A,B])$,
\begin{align}
\label{eqn:37}
G_p(\mu, \nu^{\star}) \le \frac{2}{5} - \varepsilon_p.
\end{align}

Hence,
\begin{align}
\label{eqn:38}
\inf_{\nu \in \Delta([A,B])} \sup_{\mu \in \Delta([A,B])} G_p(\mu,\nu) = \frac{2}{5} - \varepsilon_p.
\end{align}

\vskip 0.5cm

We now proceed by reasoning in the reverse direction.

\begin{enumerate}
  \item If $\operatorname{supp}(\mu) \subseteq [E,B]$, then player~2 can choose $\nu = \delta_x$ for some $x \in [A,E[$.

  \item If $\operatorname{supp}(\mu) \cap [A,E[ \neq \emptyset$, then by choosing $\nu = \delta_A$ or $\nu = \delta_{\check{A}_1}$, player~2 guarantees at least~$\frac{1}{2}$. The worst case for player~2 occurs when $\operatorname{supp}(\mu) = [A,\check{D}_1]$. Partitioning $[A,\check{D}_1]$ into the same five subintervals as above, player~2 can secure wins in at least three out of five subintervals by selecting:
  \begin{enumerate}
    \item $\nu = \delta_{\check{D}_1}$: wins on $[\check{A}_1,\check{C}_2[$, $[\check{C}_2,\check{A}_2[$, and $[\check{A}_2,\check{D}_1[$.
    \item $\nu = \delta_{\check{C}_1}$: wins on $[A,\check{C}_1[$, $[\check{C}_2,\check{A}_2[$, and $[\check{A}_2,\check{D}_1[$.
    \item $\nu = \delta_{\check{A}_1}$: wins on $[A,\check{C}_1[$, $[\check{C}_1,\check{A}_1[$, and $[\check{A}_2,\check{D}_1[$.
    \item $\nu = \delta_{\check{C}_2}$: wins on $[h_{2,p}(\check{C}_2),\check{C}_1[\,\subset\,[A,\check{C}_1[$, $[\check{C}_1,\check{A}_1[$, and $[\check{A}_1,\check{C}_2[$.
    \item $\nu = \delta_{\check{A}_2}$: wins on $[\check{C}_1,\check{A}_1[$, $[\check{A}_1,\check{C}_2[$, and $[\check{C}_2,\check{A}_2[$.
  \end{enumerate}

  Therefore, the worst case for player~2 guarantees at least~$\frac{3}{5} + \varepsilon_p$ for some $\varepsilon_p > 0$, so the corresponding payoff for player~1 is at most~$\frac{2}{5} - \varepsilon_p$.
\end{enumerate}

Hence,
\begin{align}
\label{eqn:39}
\inf_{\nu \in \Delta([A,B])} G_p(\mu,\nu) \le \frac{2}{5} - \varepsilon_p.
\end{align}

Now, define
\begin{align}
\label{eqn:40}
\mu^{\star} = 
\Big(
  \hat{p}_1\,U([A, \check{C}_1[) 
+ \hat{p}_2\,U([\check{C}_1,\check{A}_1[) 
+ \hat{p}_1\,U([\check{A}_1, \check{C}_2[) 
+ \hat{p}_2\,U([\check{C}_2, \check{A}_2[) 
+ \hat{p}_2\,U([\check{A}_2, \check{D}_1[)
\Big) \mathcal{L}.
\end{align}

Then, for all $\nu \in \Delta([A,B])$,
\begin{align}
\label{eqn:41}
G_p(\mu^{\star}, \nu) \ge \frac{2}{5} - \varepsilon_p > \frac{2}{5}.
\end{align}

This shows that
\begin{align}
\label{eqn:42}
\sup_{\mu \in \Delta([A,B])} \inf_{\nu \in \Delta([A,B])} G_p(\mu,\nu) = \frac{2}{5} - \varepsilon_p.
\end{align}

Therefore, we have established that the game has value $\frac{2}{5} - \varepsilon_p$ for some $\varepsilon_p > 0$.

\vskip 0.5cm

\subsubsection{The Critical Regime \texorpdfstring{$p = \frac{1}{6}(5 - \sqrt{13})$}{p = (1/6)(5 - sqrt(13))}}

\begin{enumerate}
  \item If $\operatorname{supp}(\nu) \subseteq [E,B]$, then player~1 can choose $\mu = \delta_x$ for some $x \in [A,E[$.

  \item If $\operatorname{supp}(\nu) \cap [A,E[ \neq \emptyset$, then by playing $\mu = \delta_A$, $\mu = \delta_{\check{A}_1}$, or $\mu = \delta_{\check{A}_2}$, player~1 guarantees at least~$\frac{1}{3}$. More specifically, player~1 can:
  \begin{enumerate}
    \item win on $[\hat{A}_2,\,\check{A}_3[$ by choosing $\mu = \delta_{\check{A}_3}$,
    \item win on $[A,\,\check{A}_1[$ by choosing $\mu = \delta_{\check{A}_1}$,
    \item win on $[\check{A}_1,\,\check{A}_2]$ by choosing $\mu = \delta_{\check{A}_2}$.
  \end{enumerate}
\end{enumerate}

Hence,
\begin{align}
\label{eqn:43}
\sup_{\mu \in \Delta([A,B])} G_p(\mu,\nu) \ge \frac{1}{3}.
\end{align}

Define
\begin{align}
\label{eqn:44}
\nu^{\star} = \frac{1}{3} 
\Big(
  U([A,\,\check{A}_1[) + U([\check{A}_1,\,\check{A}_2[) + U([\check{A}_2,\,\check{A}_3[)
\Big)\mathcal{L}.
\end{align}

Then, for all $\mu \in \Delta([A,B])$,
\begin{align}
\label{eqn:45}
G_p(\mu,\, \nu^{\star}) \le \frac{1}{3}.
\end{align}

We conclude that
\begin{align}
\label{eqn:46}
\inf_{\nu \in \Delta([A,B])} \sup_{\mu \in \Delta([A,B])} G_p(\mu,\nu) = \frac{1}{3}.
\end{align}

\vskip 0.5cm

We now consider the $\sup \inf$:

\begin{enumerate}
  \item If $\operatorname{supp}(\mu) \subseteq [E,B]$, then player~2 can choose $\nu = \delta_x$ for some $x \in [A,E[$.

  \item If $\operatorname{supp}(\mu) \cap [A,E[ \neq \emptyset$, then by choosing $\nu = \delta_A$, $\nu = \delta_{\check{A}_1}$, or $\nu = \delta_{\check{A}_2}$, player~2 guarantees at least~$\frac{2}{3}$. More precisely, player~2 can:
  \begin{enumerate}
    \item win on $[\check{A}_1,\,\check{A}_3[$ by choosing $\nu = \delta_{\check{A}_3}$,
    \item win on $[A,\,\check{A}_1[$ and $[\check{A}_2,\,\check{A}_3[$ by choosing $\nu = \delta_{\check{A}_1}$,
    \item win on $[A,\,\check{A}_2[$ by choosing $\nu = \delta_{\check{A}_2}$.
  \end{enumerate}

  Therefore, the worst-case payoff for player~1 is at most~$\frac{1}{3}$.
\end{enumerate}

Hence,
\begin{align}
\label{eqn:47}
\inf_{\nu \in \Delta([A,B])} G_p(\mu,\nu) \le \frac{1}{3}.
\end{align}

Now, define
\begin{align}
\label{eqn:48}
\mu^{\star} = \frac{1}{3} 
\Big(
  U([A,\,\check{A}_1[) + U([\check{A}_1,\,\check{A}_2[) + U([\check{A}_2,\,\check{A}_3[)
\Big)\mathcal{L}.
\end{align}

Then, for all $\nu \in \Delta([A,B])$,
\begin{align}
\label{eqn:49}
G_p(\mu^{\star},\, \nu) \ge \frac{1}{3}.
\end{align}

This shows that
\begin{align}
\label{eqn:50}
\sup_{\mu \in \Delta([A,B])} \inf_{\nu \in \Delta([A,B])} G_p(\mu,\nu) = \frac{1}{3}.
\end{align}

Therefore, the value of the game in this case is~$\frac{1}{3}$.

\vskip 0.5cm

\subsubsection{The Low-$p$ Regime \texorpdfstring{$0 < p < \frac{1}{6}(5 - \sqrt{13})$}{0 < p < (1/6)(5 - sqrt(13))}}

\hskip 0.5cm Define
\begin{align*}
m = \arg\max_{\,\{\,i \in \mathbb{N} :\, \check{A}_{2+i} \le \check{D}_1\,\}}\, i,
\quad \text{where} \quad m \ge 1.
\end{align*}

Let's calculate the $\inf \sup$}:

\begin{enumerate}
  \item If $\operatorname{supp}(\nu) \subseteq [E,B]$, then player~1 can choose $\mu = \delta_x$ for some $x \in [A,E[$.

  \item If $\operatorname{supp}(\nu) \cap [A,E[ \neq \emptyset$, then by playing $\mu = \delta_{\check{A}_i}$ for $1 \le i \le m+2$ or $\mu = \delta_{\check{D}_1}$, player~1 guarantees at least~$\frac{1}{m+2}$. Specifically:
  \begin{enumerate}
    \item player~1 wins on $[\hat{A}_1,\check{A}_{m+2}[ \subset [\check{A}_{m+1},\check{A}_{m+2}[$ and $[\check{A}_{m+2},\check{D}_1[$ by choosing $\mu = \delta_{\check{D}_1}$,
    \item player~1 wins on $[\check{A}_i,\,\check{A}_{i+1}[$ by choosing $\mu = \delta_{\check{A}_{i+1}}$ for $0 \le i \le m+1$.
  \end{enumerate}
\end{enumerate}

Hence,
\begin{align}
\label{eqn:51}
\sup_{\mu \in \Delta([A,B])} G_p(\mu,\nu) \ge \frac{1}{m+2} + \varepsilon_p.
\end{align}

Define
\begin{align}
\label{eqn:52}
\nu^{\star} =
\Big(
  \hat{p}_1\, \sum_{\substack{i=0 \\ i\,\text{odd}}}^{m+1}
  U([\check{A}_{i},\, \hat{D}_{m+1+i}[)
  + \hat{p}_2\, \sum_{\substack{i=0 \\ i\,\text{even}}}^{m+1}
  U([\hat{D}_{m+1+i},\, \check{A}_{i+1}[)
  + \hat{p}_1\, U([\check{A}_{m+2},\, \check{D}_1[)
\Big)\mathcal{L}.
\end{align}

Then, for all $\mu \in \Delta([A,B])$,
\begin{align}
\label{eqn:53}
G_p(\mu,\, \nu^{\star}) \le \frac{1}{m+2} + \varepsilon_p.
\end{align}

This shows
\begin{align}
\label{eqn:54}
\inf_{\nu \in \Delta([A,B])} \sup_{\mu \in \Delta([A,B])} G_p(\mu,\nu)
= \frac{1}{m+2} + \varepsilon_p.
\end{align}

\vskip 0.5cm

We calculate now the $\sup \inf$:

\begin{enumerate}
  \item If $\operatorname{supp}(\mu) \subseteq [E,B]$, then player~2 can choose $\nu = \delta_x$ for some $x \in [A,E[$.

  \item If $\operatorname{supp}(\mu) \cap [A,E[ \neq \emptyset$, then by playing $\nu = \delta_{\check{A}_i}$ for $1 \le i \le m+2$ or $\nu = \delta_{\check{D}_1}$, player~2 guarantees at most~$\frac{m+1}{m+2}$. Specifically:
  \begin{enumerate}
    \item player~2 wins on $[\check{A}_1,\check{D}_1[$ by choosing $\nu = \delta_{\check{D}_1}$,
    \item player~2 wins on $[\check{A}_1,\check{D}_1[\, \setminus\, [\check{A}_{i-1},\check{A}_i[$ by choosing $\nu = \delta_{\check{A}_i}$ for $0 \le i \le m+1$,
    \item player~2 wins on $[\check{A}_1,\check{D}_1[\, \setminus\, \big([A,\, h_{2,p}(\check{A}_{m+2})[\, \cup\, [\check{A}_{m+2},\,\check{D}_1[\,\big)$ by choosing $\nu = \delta_{\check{A}_{m+2}}$.
  \end{enumerate}
\end{enumerate}

Hence,
\begin{align}
\label{eqn:55}
\inf_{\nu \in \Delta([A,B])} G_p(\mu,\nu) \le \frac{1}{m+2} + \varepsilon_p.
\end{align}

Define
\begin{align}
\label{eqn:56}
\mu^{\star} =
\Big(
  \hat{p}_1\, \sum_{\substack{i=0 \\ i\,\text{odd}}}^{m+1}
  U([\check{A}_{i},\, \hat{D}_{m+1+i}[)
  + \hat{p}_2\, \sum_{\substack{i=0 \\ i\,\text{even}}}^{m+1}
  U([\hat{D}_{m+1+i},\, \check{A}_{i+1}[)
  + \hat{p}_1\, U([\check{A}_{m+2},\, \check{D}_1[)
\Big)\mathcal{L}.
\end{align}

For all $\nu \in \Delta([A,B])$,
\begin{align}
\label{eqn:57}
G_p(\mu^{\star},\, \nu) \ge \frac{1}{m+2} + \varepsilon_p.
\end{align}

This implies
\begin{align}
\label{eqn:58}
\sup_{\mu \in \Delta([A,B])} \inf_{\nu \in \Delta([A,B])} G_p(\mu,\nu)
= \frac{1}{m+2} + \varepsilon_p.
\end{align}

Therefore, the game has value~$\frac{1}{m+2} + \varepsilon_p$ for some~$\varepsilon_p > 0$.

\vskip 0.5cm

\subsection{Explicit Solution}

\hskip 0.5cm One limitation of the previous section's analysis is the lack of an explicit closed form for the game value, particularly the term~$\varepsilon_p$, which we compute numerically. In this section, we present an alternative approach to obtain an explicit expression for the value using a functional system.\\

Suppose that players~1 and~2 adopt mixed strategies with densities~$f_1$ and~$f_2$, respectively. Integrating~$g_p$ with respect to~$f_2$ gives
\begin{align}
\label{eqn:59}
\begin{split}
\int g_p(x,y)\, f_2(y)\, dy
&= \mathbf{1}_{\{\, x < E \}} \!\Big(\int_{\frac{p\,x + E}{p+1}}^{E} f_2(y)\, dy\Big)
+ \mathbf{1}_{\{\, x < E \}} \!\Big(\int_{\frac{(p+1)x - E}{p}}^{x} f_2(y)\, dy\Big)\\
&\quad + \mathbf{1}_{\{\, E \le x \}} \!\Big(\int_{x}^{\frac{(p+1)x - E}{p}} f_2(y)\, dy\Big)
+ \mathbf{1}_{\{\, E \le x \}} \!\Big(\int_{\frac{(p+1)x - E}{p}}^{B} f_2(y)\, dy\Big).
\end{split}
\end{align}

Applying the same reasoning as in Section~\ref{The functional equation solution}, for $x$ in $\operatorname{supp}(f_1)$, we have
\begin{align}
\label{eqn:60}
\begin{split}
\int_{\frac{p\,x + E}{p+1}}^{E} f_2(y)\, dy + \int_{A}^{x} f_2(y)\, dy &= v(p), \quad A \le x < \hat{A}_1,\\
\int_{\frac{p\,x + E}{p+1}}^{E} f_2(y)\, dy + \int_{\frac{(p+1)x - E}{p}}^{x} f_2(y)\, dy &= v(p), \quad \hat{A}_1 \le x < E,\\
\int_{x}^{B} f_2(y)\, dy &= v(p), \quad E \le x \le B.
\end{split}
\end{align}

Differentiating with respect to~$x$ yields
\begin{align}
\label{eqn:61}
\begin{split}
-\,\frac{p}{p+1}\, f_2\!\Big(\tfrac{p\,x + E}{p+1}\Big) + f_2(x) &= 0, \quad A \le x < \hat{A}_1,\\
-\,\frac{p}{p+1}\, f_2\!\Big(\tfrac{p\,x + E}{p+1}\Big) + f_2(x) - \frac{p+1}{p}\, f_2\!\Big(\tfrac{(p+1)x - E}{p}\Big) &= 0, \quad \hat{A}_1 \le x < E,\\
-\,f_2(x) &= 0, \quad E \le x \le B.
\end{split}
\end{align}

The last condition shows that $f_2$ vanishes on~$[E,B]$. Solving on~$[A,\,\hat{A}_1[$
\begin{align}
\label{eqn:62}
f_2(x) = \frac{p}{p+1}\, f_2\!\Big(\tfrac{p\,x + E}{p+1}\Big),
\end{align}
whose solution is
\begin{align}
\label{eqn:63}
f_2(x) = C_2\, \frac{1}{x - E}.
\end{align}

A similar argument for player~1 gives
\begin{align}
\label{eqn:64}
\begin{split}
\int g_p(x,y)\, f_1(x)\, dx
&= \mathbf{1}_{\{\, y < E \}} \!\Big(\int_{y}^{\frac{(1-p)\,y + E}{2 - p}} f_1(x)\, dx\Big)
+ \mathbf{1}_{\{\, y < E \}} \!\Big(\int_{A}^{\frac{(1+p)y - E}{p}} f_1(x)\, dx\Big)\\
&\quad + \mathbf{1}_{\{\, E \le y \}} \!\Big(\int_{\frac{(1-p)\,y + E}{2 - p}}^{y} f_1(x)\, dx\Big)
+ \mathbf{1}_{\{\, E \le y \}} \!\Big(\int_{E}^{\frac{(1-p)\,y + E}{2 - p}} f_1(x)\, dx\Big).
\end{split}
\end{align}

Then, for $y$ in $\operatorname{supp}(f_2)$
\begin{align}
\label{eqn:65}
\begin{split}
\int_{y}^{\frac{(1-p)\,y + E}{2 - p}} f_1(x)\, dx &= v(p), \quad A \le y < \hat{A}_1,\\
\int_{y}^{\frac{(1-p)\,y + E}{2 - p}} f_1(x)\, dx + \int_{A}^{\frac{(1+p)y - E}{p}} f_1(x)\, dx &= v(p), \quad \hat{A}_1 \le y < E,\\
\int_{E}^{y} f_1(x)\, dx &= v(p), \quad E \le y \le B.
\end{split}
\end{align}

Differentiating gives
\begin{align}
\label{eqn:66}
\begin{split}
\frac{1 - p}{2 - p}\, f_1\!\Big(\tfrac{(1 - p)\,y + E}{2 - p}\Big) - f_1(y) &= 0, \quad A \le y < \hat{A}_1,\\
\frac{1 - p}{2 - p}\, f_1\!\Big(\tfrac{(1 - p)\,y + E}{2 - p}\Big) - f_1(y) + \frac{1 + p}{p}\, f_1\!\Big(\tfrac{(1 + p)\,y - E}{p}\Big) &= 0, \quad \hat{A}_1 \le y < E,\\
f_1(y) &= 0, \quad E \le y \le B.
\end{split}
\end{align}

Thus, $f_1$ is also zero on~$[E,B]$. Solving on~$[A,\,\hat{A}_1[$
\begin{align}
\label{eqn:67}
f_1(y) = \frac{1 - p}{2 - p}\, f_1\!\Big(\tfrac{(1 - p)\,y + E}{2 - p}\Big),
\end{align}
which implies
\begin{align}
\label{eqn:68}
f_1(x) = C_1\, \frac{1}{x - E}.
\end{align}

Imposing that the supports of $f_1$ and $f_2$ are~$[A,\,\tilde{A}]$ with
\begin{align}
\label{eqn:69}
\tilde{A} = \check{D}_1 = \frac{\, 2E + p(1 - p)\, A\,}{(2 - p)(p + 1)},
\end{align}
as determined in the previous analysis, and enforcing
\begin{align}
\label{eqn:70}
\int_{A}^{\tilde{A}} f_1(x)\, dx = 1, 
\quad \int_{A}^{\tilde{A}} f_2(y)\, dy = 1,
\end{align}
we find
\begin{align}
\label{eqn:71}
C_1 = C_2 = \Big[\ln\!\Big( \frac{E - \tilde{A}}{E - A} \Big)\Big]^{-1}.
\end{align}

Finally, integrating $g_p$ yields the explicit game value
\begin{align}
\label{eqn:72}
\begin{split}
v(p)
&= \iint_{[A,\tilde{A}]^2} g_p(x,y)\, f_1(x)\, f_2(y)\, dx\, dy\\
&= \frac{\ln \left(\frac{1-p}{2-p}\right)}{\ln \left(\frac{p(1-p)}{(2-p)(p+1)}\right)}.
\end{split}
\end{align}

One can check that by taking
\begin{align}
\label{eqn:73}
f^{\star}(x) &= \frac{1}{\Big[\ln\!\Big( \frac{E - \tilde{A}}{E - A} \Big)\Big] (x - E)} \mathbf{1}_{[A,\tilde{A}]}.
\end{align}

\noindent that
\begin{align}
\label{eqn:74}
\begin{split}
\int_{[A,B]} g_p(x,y)\, f^{\star}(x)\, dx &= \frac{\ln \left(\frac{1-p}{2-p}\right)}{\ln \left(\frac{p(p-1)}{(p-2)(p+1)}\right)}  \, \mathbf{1}_{[A,\tilde{A}]} + \frac{\ln \left(\frac{(p+1) (E-y)}{p (E-A)}\right)}{\ln \left(\frac{p(p-1)}{(p-2)(p+1)}\right)} \mathbf{1}_{[\tilde{A},\tilde{A}_2[} + \mathbf{1}_{[\tilde{A}_2,B]} .
\end{split}
\end{align}

\noindent and
\begin{align}
\label{eqn:75}
\begin{split}
\int_{[A,B]} g_p(x,y)\, f^{\star}(y)\, dy &= \frac{\ln \left(\frac{1-p}{2-p}\right)}{\ln \left(\frac{p(p-1)}{(p-2)(p+1)}\right)}  \, \mathbf{1}_{[A,\tilde{A}]} + \frac{\ln \left(\frac{p (p-1)^2 (E-A)}{(p-2)^2 (p+1) (E-x)}\right)}{\ln
   \left(\frac{p (p-1)}{(p-2)(p+1)}\right)} \mathbf{1}_{[\tilde{A},\tilde{A}_3[} .
\end{split}
\end{align}

\noindent where $\tilde{A}_2=\frac{A (p-1) p^2+ E(p(p-3)-2)}{(p-2) (p+1)^2}$ and $\tilde{A}_3=\frac{A \, p(p-1)^2 - E \left(p^2+p-4\right)}{(p-2)^2 (p+1)}$.

\vskip 0.5cm

\section{The $N$-players version\label{The $N$-players version}}

\hskip 0.5cm When $N$ players ($N \ge 3$) participate in the Moroccan public procurement game, the situation becomes considerably more complex. To generalize the analysis of Section~\ref{The two players version} to the $N$-player case, let us denote
\begin{align}
\label{eqn:76}
N^-=\{ 1\leq j \leq N \, \mid \, x_j < P \} \quad , \quad N^0=\{ 1\leq j \leq N \, \mid \, x_j = P \} \quad , \quad N^+=\{ 1\leq j \leq N \, \mid \, x_j > P \}
\end{align}

The payoff function of player $i$ is given by the discontinuous function
\begin{align}
\label{eqn:77}
g_i(x_1,\hdots,x_N) &= \sum_{n=0}^{N-1} \sum_{J_n(i)} \frac{1}{n+1} \prod_{k \in J_n(i)} \mathbf{1}_{\{x_i = x_k\}} \left( \mathbf{1}_{\{ x_i \leq P \}} \prod_{j\neq i, \, j \in N^- \setminus J_n(i)} \mathbf{1}_{\{x_j < x_i\}} + \mathbf{1}_{\{ x_i > P \}} \prod_{j\neq i, \, j \not \in J_n(i)} \mathbf{1}_{\{ x_i < x_j \}}  \right) 
\end{align}

\noindent where $J_n(i)$ is a subset of $n$ elements of $\{1,\hdots,i-1,i+1,\hdots,N\}$ and
\begin{align}
\label{eqn:78}
P(x_1,\hdots,x_N)&=\frac{\sum_{i=1}^N x_i + N\,E}{2N}
\end{align}

The situation resembles a class of aggregative games, with the additional feature that the payoff functions depend not only on the player's own strategy and the aggregator, but also on the strategies of the other players. Moreover, the direct approach fails in this setting for $N \ge 3$, making functional derivation of the solution impracticable.

\vskip 0.5cm

\subsection{Illustration of the case $N=3$}

\begin{figure}[!htb]
\begin{center}
\begin{minipage}[t]{0.3\textwidth}
  \centering
  \includegraphics[width=\linewidth]{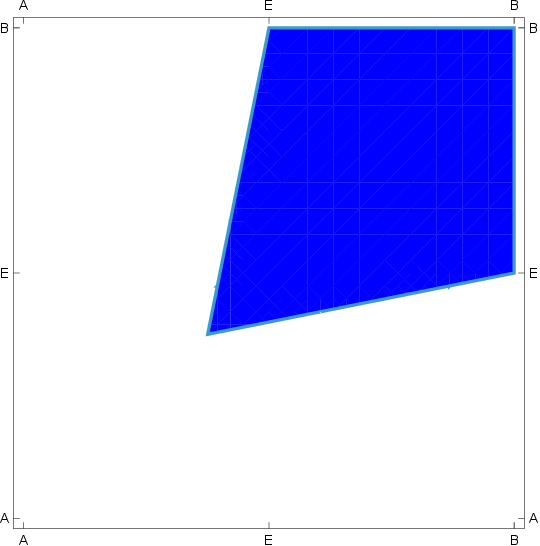}
  \caption*{$x=A$.}
\end{minipage}
\hfill
\begin{minipage}[t]{0.3\textwidth}
  \centering
  \includegraphics[width=\linewidth]{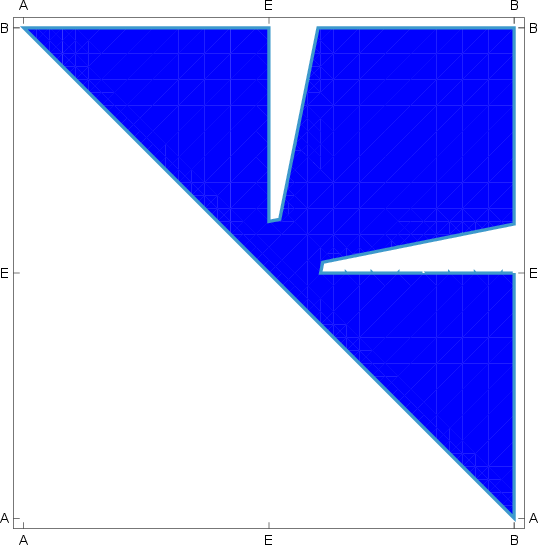}
  \caption*{$x=E$.}
\end{minipage}
\hfill
\begin{minipage}[t]{0.3\textwidth}
  \centering
  \includegraphics[width=\linewidth]{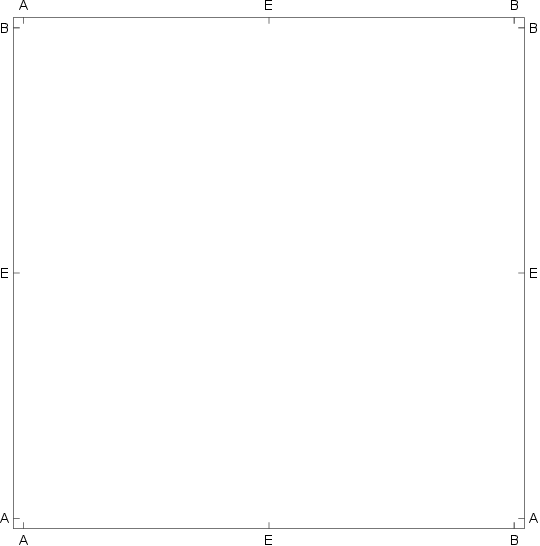}
  \caption*{$x=B$.}
\end{minipage}
\end{center}
\captionof{figure}{The winning region for different values of $x$.}
\label{Graph3}
\end{figure}

\hskip 0.5cm Let us illustrate the specific case of three players. In this setting, the payoff function of player~1 is given by
\begin{align}
\label{eqn:79}
\begin{split}
g(x,y,z)&=  \mathbf{1}_{\{ z \le y < x \le P \}} + \mathbf{1}_{\{ y < z < x \le P \}} + \mathbf{1}_{\{ y < x \le P < z \}} + \mathbf{1}_{\{ z < x \le P < y \}} \\
&+ \mathbf{1}_{\{x \le P < z \le y \}} + \mathbf{1}_{\{x \le P < y < z \}} + \mathbf{1}_{\{P < x < z \le y \}} + \mathbf{1}_{\{P < x < y < z \}} \\
&+  \frac{1}{2} \left( \mathbf{1}_{\{ z < y = x \le P \}} + \mathbf{1}_{\{ y < z = x \le P \}} + \mathbf{1}_{\{ y = x \le P < z\}} + \mathbf{1}_{\{ z = x \le P < y\}} + \mathbf{1}_{\{ P \le y = x < z\}} + \mathbf{1}_{\{P \le z = x < y\}} \right) \\
&+ \frac{1}{3} \mathbf{1}_{\{x=y=z\}}
\end{split}
\end{align}

\noindent given the bids $x$, $y$, and $z$ of the three players, respectively. We can represent the winning region ($g(x,y,z)=1$) of player~1 for different values of $x$ (see Figure~\ref{Graph3}):

\vskip 0.5cm

\vskip 0.5cm

We will show that $g$ has a finite number of jump points and can be expressed accordingly. Rearranging the inequality $x \le P$ gives
\begin{align*}
x \le \frac{x+y+z+3E}{6} \iff x \le \frac{y+z+3E}{5}.
\end{align*}
Define the opponents-only threshold
\begin{align}
\label{eqn:80}
t(y,z):=\frac{y+z+3E}{5}.
\end{align}

For fixed $(y,z)$, ties of the form $x=t(y,z)=y$ or $x=t(y,z)=z$ introduce two additional $x$-values at which the ordering of $\{x,y,z,t(y,z)\}$ changes. Solving
\begin{align}
\label{eqn:81}
P(x,y,z)=y\iff x=5y-3E-z=:p_y(y,z),
\quad
P(x,y,z)=z\iff x=5z-3E-y=:p_z(y,z),
\end{align}

We obtain two more opponent-dependent cutpoints. Thus for fixed $(y,z)$ the function $x\mapsto g(x,y,z)$ depends only on the ordering of $\{\,y,\ z,\ t(y,z),\ p_y(y,z),\ p_z(y,z)\,\}$ and is piecewise constant with possible discontinuities only at these five points.\\

Fix $(y,z)$. For each $s\in\mathcal S(y,z):=\{\,y,\ z,\ t(y,z),\ p_y(y,z),\ p_z(y,z)\,\}$ define the one-sided interval limits
\begin{align*}
g_1(s^-,y,z):=\lim_{x\uparrow s} g_1(x,y,z),\qquad
g_1(s^+,y,z):=\lim_{x\downarrow s} g_1(x,y,z).
\end{align*}
These one-sided limits are values taken on open intervals adjoining $s$, hence belong to $\{0,1\}$. Define the interval-jump
\begin{align}
\label{eqn:82}
\Delta_s(y,z):=g_1(s^+,y,z)-g_1(s^-,y,z)\in\{-1,0,1\}.
\end{align}

The five cutpoints satisfy the linear relations
\begin{align*}
p_y-y = 5(y-t),\qquad
p_z-z = 5(z-t),\qquad
p_y-p_z = 6(y-z).
\tag{S}
\end{align*}

Hence $p_y$ lies on the same side of $t$ as $y$, $p_z$ lies on the same side of $t$ as $z$, and $p_y<p_z\iff y<z$.  
Only four geometric orderings (up to symmetry \(y\leftrightarrow z\)) are compatible with (S):

\medskip
\noindent\textbf{Case I: \(y<t\) and \(z<t\).}
Two orderings are possible:
\begin{align*}
(O_1)\quad p_y<p_z<y<z<t,
\qquad
(O_2)\quad p_y<y<p_z<z<t.
\end{align*}

\noindent\textbf{Case II: \(y<t<z\).}
A single ordering is possible:
\begin{align*}
(O_3)\quad p_y<y<t<z<p_z.
\end{align*}

\noindent\textbf{Case III: \(t<y<z\).}
Two orderings are possible:
\begin{align*}
(O_4)\quad t<y<z<p_y<p_z.
\qquad
(O_5)\quad t<y<p_y<z<p_z.
\end{align*}

\medskip
\noindent These are the \emph{only} admissible relative orders of $\{p_y,p_z,y,z,t\}$.\\

\vskip 0.5cm

\begin{figure}[!htb]
\begin{center}
\includegraphics[scale=1]{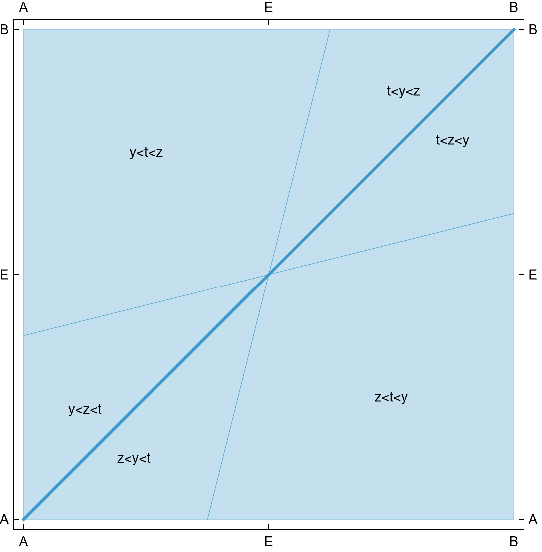}
\end{center}
\captionof{figure}{The six regions of the $y-z$ plane.}
\label{Graph4}
\end{figure}

\vskip 0.5cm

We can summarize the sign of $\Delta_s$ in the following table:

\vspace{0.5cm}

\begin{table}[!htb]
\centering
\renewcommand{\arraystretch}{1.25}

\begin{tabular}{c|ccccc}
\hline
\multicolumn{1}{c|}{$O_k$ / $s$} 
  & $y$ & $p_y$ & $z$ & $p_z$ & $t$ \\
\hline
$O_1$ 
  & \phantom{-}0 & -1 & +1 & \phantom{-}0 & -1 \\[0.2em]
$O_2$ 
  & +1 & -1 & +1 & -1 & -1 \\[0.2em]
$O_3$ 
  & +1 & -1 & \phantom{-}0 & \phantom{-}0 & -1 \\[0.2em]
$O_4$ 
  & -1 & \phantom{-}0 & \phantom{-}0 & \phantom{-}0 & \phantom{-}0 \\[0.2em]
$O_5$ 
  & -1 & \phantom{-}0 & \phantom{-}0 & \phantom{-}0 & \phantom{-}0 \\
\hline
\end{tabular}

\caption{Interval-jump signs $\Delta_s$ for each ordering cell $O_k$.}
\label{tab:deltas}
\end{table}

\vspace{0.5cm}

\subsection{Existence result for $N \geq 3$}

\hskip 0.5cm Let's start by noticing that the game is symmetric, since 
\begin{align}
\label{eqn:83}
g_{\pi(i)}(x_{\pi(1)},\dots,x_{\pi(N)}) &= g_i(x_1,\dots,x_N)
\end{align}
for any permutation $\pi$ and the set of actions is $[A,B]$ for all $i$. Moreover, the function $g_i$ is measurable and bounded between $0$ and $1$, but it is neither quasiconcave nor continuous. The strategy set, is convex and compact.\\

Next, we study the set of discontinuities of $g_i$, denoted by $\mathcal{D}_i$. It is formed by the following disjoints hypersurfaces:

\begin{enumerate}
\item \textbf{Hypersurfaces of ties:}
    \begin{align}
    \label{eqn:84}
    \mathcal{T}_i = \left\lbrace (x_1,\dots,x_N) \in [A,B]^N \;\middle|\; \exists n, \, 1 \le n \le N, \, x_i = x_k \text{ for } k \in J_n(i), \text{ and } g_i(x_1,\dots,x_N) = \frac{1}{n} \right\rbrace
    \end{align}

\item \textbf{Hypersurfaces of fixed points:}
    \begin{align}
    \label{eqn:85}
    \mathcal{F}_i = \left\lbrace (x_1,\dots,x_N) \in [A,B]^N \;\middle|\; x_i = \frac{\sum_{j\neq i} x_j + N E}{2N-1}, \quad g_i(x_1,\dots,x_N) = 1 \right\rbrace
    \end{align}
    
\item \textbf{Hypersurfaces of transition points:} define $\displaystyle \underline{j}\in \text{Arg}\max_{1\leq j\leq N} \left\{ x_j \, \middle| \, x_j \leq \frac{\sum_{j\neq i} x_j + N E}{2N-1} \right\}$
    \begin{align}
    \label{eqn:86}
    \mathcal{P}_i = \left\lbrace (x_1,\dots,x_N) \in [A,B]^N \;\middle|\; x_i = (2N-1)x_{\underline{j}} - \sum_{j\neq i,\underline{j}} x_j - N E, \quad g_i(x_1,\dots,x_N) = 0 \right\rbrace
    \end{align}
\end{enumerate}

The set $\mathcal{D}_i$ is closed and has zero $N$-dimensional Lebesgue measure, i.e., $\mathcal{L}^N(\mathcal{D}_i) = 0$.\\

The function $g_i$ is neither lower semi-continuous nor upper semi-continuous. Counterexamples can be found at diagonal points $(x,\dots,x) \in \mathcal{T}_i$ for $x \in [A,B]$.\\

The function $g_i$ is, however, upper semi-continuous on $[A,B]^N \setminus \mathcal{T}_i\bigcup \mathcal{P}_i$. Indeed, for all $i \in \{1,\dots,N\}$, for any $\mathbf{x} = (x_1,\dots,x_N) \in \mathcal{F}_i$, and for any sequence $(\mathbf{x}_n)$ such that $\lim_{n \to \infty} \mathbf{x}_n = \mathbf{x}$, we have
\begin{align}
\label{eqn:87}
\lim_{n \to \infty} \sup g_i(\mathbf{x}_n)  \le g_i(\mathbf{x}) = 1.
\end{align}

One can verify that $g_i(x_i,\mathbf{x}_{-i})$ is weakly lower semi-continuous in $x_i$ for all $i$ in the sens of~\cite{Dasgupta1986}:

\vskip 0.5cm

\begin{definition}{\textbf{Weak lower semicontinuity~\cite{Dasgupta1986}}}\\
Let $g_i : [A,B]^N \to \R$ denote player $i$'s payoff function. We say that $g_i(x_i,\mathbf{x}_{-i})$ is \emph{weakly lower semicontinuous} in $x_i$ if, for every $z_i \in \mathcal{D}_i$, there exists $\lambda \in [0,1]$ such that, for every $\mathbf{x}_{-i} \in \mathcal{D}_{-i}(z_i)$
\begin{align*}
g_i(z_i,\mathbf{x}_{-i})
\;\le\;
\lambda \liminf_{x_i \to \overline{x}_i^-} g_i(z,\hdots,z)
\;+\;
(1-\lambda) \liminf_{x_i \to \overline{x}_i^+} g_i(z,\hdots,z).
\end{align*}
\end{definition}

\vskip 0.5cm

Indeed, for $\mathbf{x} \in \mathcal{T}_i$ for some $i$, $g_i$ is easily seen to be either left- or right-lower semi-continuous. If $\mathbf{x} \in \mathcal{F}_i$ (the single-winner situation), the same reasoning applies. If $\mathbf{x} \in \mathcal{P}_i$, one can check that the function is lower semi-continuous: for any sequence $(\mathbf{x}_n)$ such that $\lim_{n \to \infty} \mathbf{x}_n = \mathbf{x}$, we have
\begin{align}
\label{eqn:88}
\lim_{n \to \infty} \inf g_i(\mathbf{x}_n) \ge g_i(\mathbf{x})=0.
\end{align}

Moreover, the sum $\sum_{i=1}^N g_i$ is upper semi-continuous, since the sum always equals $1$. Unfortunately, the discontinuity set cannot be defined by a one-to-one correspondence, and Lemma 3 of~\cite{Dasgupta1986} fails to apply as soon as $N \ge 3$.\\

The upper semi-continuity of the game implies its reciprocal upper semi-continuity introduced in~\cite{Reny1999}, while one can prove that the game is payoff secure since, for all $i$ and all $\mathbf{x}_{-i}\in [A,B]^{N-1}$,
\begin{align*}
\sup_{x_i\in[A,B]} g_i(x_i,\mathbf{x}_{-i})=\sup_{x_i\in[A,B]} \underline{g_i}(x_i,\mathbf{x}_{-i})=1
\end{align*}

\noindent where $\underline{g_i}$ represents the lower semi-continuous regularization of $g_i$ (obtained by replacing its values on $\mathcal{T}_i$ and $\mathcal{F}_i$ with zero). Indeed, one can use the alternative definition of payoff security (\cite{Reny1999}) to prove the following: For all $\mathbf{x}_{-i} \in S_{-i}$ define
\begin{align}
\label{eqn:89}
t(\mathbf{x}_{-i})&=\frac{\sum_{j\neq i} x_j + N E}{2N-1}
\end{align}

We distinguish two cases:
\begin{enumerate}
\item $\not \exists j \, : \, x_j =t(\mathbf{x}_{-i})$: Let  $\displaystyle \underline{j}=\arg\max_{1\leq j\leq N} \left\{ x_j \, \middle| \, x_j < t(\mathbf{x}_{-i}) \right\}$ and  $\displaystyle \overline{j}=\arg\min_{1\leq j\leq N} \left\{ x_j \, \middle| \, x_j > t(\mathbf{x}_{-i}) \right\}$, then define
\begin{align*}
\underline{\varepsilon} &= \mid t(\mathbf{x}_{-i})  - x_{\underline{j}}  \mid \\
\overline{\varepsilon} &= \mid t(\mathbf{x}_{-i})  - x_{\overline{j}}  \mid
\end{align*}

Then fix $\displaystyle \tilde{\varepsilon}=\varepsilon< \frac{\min(\underline{\varepsilon} , \overline{\varepsilon})}{2}$ and define 
\begin{align*}
\bar{x}_i &= t(\mathbf{x}_{-i}) - \varepsilon \\
N_{\tilde{\varepsilon}}(\mathbf{x}_{-i}) &= \left\{
\mathbf{y}_{-i} \in S_{-i} :
\|\mathbf{y}_{-i} - \mathbf{x}_{-i}\| < \tilde{\varepsilon} \right\}.
\end{align*}

\item $\exists j \, : \, x_j =t(\mathbf{x}_{-i})$: Let  $\displaystyle \underline{j}=\arg\max_{1\leq j\leq N} \left\{ x_j \, \middle| \, x_j < t(\mathbf{x}_{-i}) \right\}$ and  $\displaystyle \overline{j}=\arg\min_{1\leq j\leq N} \left\{ x_j \, \middle| \, x_j > t(\mathbf{x}_{-i}) \right\}$, then define
\begin{align*}
\underline{\varepsilon} &= \mid t(\mathbf{x}_{-i})  - x_{\underline{j}}  \mid \\
\overline{\varepsilon} &= \mid t(\mathbf{x}_{-i})  - x_{\overline{j}}  \mid
\end{align*}

Then fix $\displaystyle \varepsilon< \frac{\min(\underline{\varepsilon} , \overline{\varepsilon})}{2}$ and $\tilde{\varepsilon}<\frac{\varepsilon}{2N}$ and define 
\begin{align*}
\bar{x}_i &= t(\mathbf{x}_{-i}) - \varepsilon \\
N_{\tilde{\varepsilon}}(\mathbf{x}_{-i}) &= \left\{
\mathbf{y}_{-i} \in S_{-i} :
\|\mathbf{y}_{-i} - \mathbf{x}_{-i}\| < \tilde{\varepsilon}
\right\}.
\end{align*}

\item $\forall j \, : \, x_j =t(\mathbf{x}_{-i})$: Fix $\displaystyle \varepsilon>0$ and $\tilde{\varepsilon}<\frac{\varepsilon}{2N}$ and define 
\begin{align*}
\bar{x}_i &= t(\mathbf{x}_{-i}) - \varepsilon \\
N_{\tilde{\varepsilon}}(\mathbf{x}_{-i}) &= \left\{
\mathbf{y}_{-i} \in S_{-i} :
\|\mathbf{y}_{-i} - \mathbf{x}_{-i}\| < \tilde{\varepsilon}
\right\}.
\end{align*}
\end{enumerate}

In all the cases one can check that
\begin{align}
\label{eqn:90}
\mathbf{y}_{-i} \in N_{\tilde{\varepsilon}}(\mathbf{x}_{-i}) \;\Longrightarrow\; 
g_i(\bar{x}_i, \mathbf{y}_{-i})=1 \ge g_i(x_i, \mathbf{x}_{-i}).
\end{align}

Thus, the game is better-reply secure \cite[Prop. 3.2]{Reny1999}. However, the game is not quasiconcave which prevent the application of Reny's fundamental result about equilibrium existence in pure strategies.\\

As pointed out by \cite{Reny1999}, when moving to mixed strategies, securing a payoff becomes in certain respects easier, yet also more difficult. In fact, establishing payoff security in mixed strategies is considerably harder than in the pure-strategy case. To facilitate this verification, several authors have proposed conditions on the underlying pure-strategy game that allow one to conclude about payoff security of its mixed extension. A first such condition is the notion of \emph{uniformly payoff secure games}, introduced by~\cite{Monteiro2007}.

\vskip 0.5cm

\begin{definition}{\textbf{Uniformly payoff security~(\cite{Monteiro2007})}}\\
A game $\{(g_i,S_i)_{i=1}^N\}$ is \emph{uniformly payoff secure} if for every $x_i \in S_i$ and every $\varepsilon > 0$ there exists for each player $i$ a strategy 
$\bar{x}_i \in S_i$ such that for every $\mathbf{x}_{-i} \in S_{-i}$ there exists a neighborhood $N(\mathbf{x}_{-i})$ of $\mathbf{x}_{-i}$ with the property that
\begin{align}
\label{eqn:91}
\mathbf{y}_{-i} \in N(\mathbf{x}_{-i}) \;\Longrightarrow\; 
g_i(\bar{x}_i, \mathbf{y}_{-i}) \ge g_i(x_i, \mathbf{x}_{-i}) - \varepsilon.
\end{align}
\end{definition}

\vskip 0.5cm

On can check that our game is not uniformly payoff secure. Take for example $N=2$, and choose $x_i<E$, then
\begin{enumerate}
\item If $x_i<\bar{x}_i$, choose $y$ s.t. $y<x_i<t<\bar{x}_i$.

\item If $x_i>\bar{x}_i$, choose $y$ s.t. $\bar{x}_i<y<x_i<t$.

\item If $x_i=\bar{x}_i$, choose $y=x_i=\bar{x}_i$.
\end{enumerate}

The game is not uniformly diagonally secure in the sens of~\cite{Prokopovych2014} neither.\\

A less demanding condition is introduced by \cite{Allison2014} and can check directly the payoff security in the mixed extension.

\vskip 0.5cm

\begin{definition}{\textbf{Disjoint Payoff Matching~(\cite{Allison2014})}}\\
A game $\{(g_i,S_i)_{i=1}^N\}$ satisfies \emph{disjoint payoff matching} if for all $x_i \in S_i$, there exists a sequence of deviations $(x_i^k)_{k\geq 1} \subset S_i$ such that the following hold:
\begin{align}
\label{eqn:92}
\begin{split}
&1. \qquad \liminf_{k\to\infty} g_i(x_i^k, \mathbf{x}_{-i})  \;\ge\; g_i(x_i, \mathbf{x}_{-i}) \qquad \forall\, \mathbf{x}_{-i} \in S_{-i}, \\
&2. \qquad \limsup_{k\to\infty} \mathbb{D}_i(x_i^k) = \emptyset.
\end{split}
\end{align}
\noindent where 
\begin{align}
\label{eqn:93}
\mathbb{D}_i(x_i) &= \left\{ \, \mathbf{x}_{-i} \in S_{-i} \; : \; g_i \text{ is discontinuous in } \mathbf{x}_{-i} \text{ at } (x_i, \mathbf{x}_{-i})\right\}.
\end{align}
\end{definition}

\vskip 0.5cm

\begin{theorem}{\textbf{(\cite{Allison2014})}}\\
Let $\{(g_i,S_i)_{i=1}^N\}$ be a compact game and satisfies the disjoint payoff matching condition. Then its mixed extension is payoff secure.
\end{theorem}

\vskip 0.5cm

This result, together with the upper semicontinuity of the game, ensures that the mixed extension is better-reply secure, since upper semicontinuity of the pure strategy game is inherited by its mixed extension.

\vskip 0.5cm

\begin{theorem}{\textbf{\cite[Corollary 5.2]{Reny1999}} }\\
Suppose that $\{(g_i,S_i)_{i=1}^N\}$ is a compact, Hausdorff game. Then it possesses a mixed–strategy Nash equilibrium if its mixed extension is better–reply secure. Moreover, its mixed extension is better–reply secure if it is both reciprocally upper semicontinuous and payoff secure.
\end{theorem}

\vskip 0.5cm

Let's try to construct such a sequence of deviations $(x_i^k)$ for our game $\{(g_i,[A,B])_{i=1}^N\}$. Since $\mathcal{T}_i(x_i)$, $\mathcal{F}_i(x_i)$ and $\mathcal{P}_i(x_i)$ are mutually disjoint, we define the sequence of deviations $(x_i^k)$ as follows. 

\begin{enumerate}
\item For $x_i \in ]A,B]$, choose a strictly increasing sequence $x_i^k \uparrow x_i$ in a neighborhood of $x_i$, then:

\begin{enumerate}
\item If $(x_i,\mathbf{x}_{-i})$ is a continuity point, the inferior limit condition is trivially satisfied.

\item If $\mathbf{x}_{-i} \in \mathcal{F}_i(x_i)$, one can verify that $g_i(\cdot,\mathbf{x}_{-i})$ in a neighborhood of $x_i$ decreases drastically from $1$ to $0$ for $x > x_i$.

\item If $\mathbf{x}_{-i} \in \mathcal{P}_i(x_i)$, one can verify that $g_i(\cdot,\mathbf{x}_{-i})$ in a neighborhood of $x_i$ takes the value $1$ below $x_i$ and becomes $0$ for $x \geq x_i$.
\end{enumerate}

\item For $x_i=A$, choose $x_i^k \downarrow A$.
\end{enumerate}

In addition to the fact that the maps $s_1(x_j)=x_j$, $s_2(\mathbf{x}_{-i})=\sum_{j\neq i} x_j$ and $s_3(\mathbf{x}_{-i})=(2N-1)x_{\underline{j}} - \sum_{j\neq i,\underline{j}} x_j$ are single–valued functions, we conclude that $\mathcal{D}_i(x_i^k) \cap \mathcal{D}_i(x_i^l) = \emptyset$ for $k \neq l$. However, we get a problem at ties situation: If $\mathbf{x}_{-i} \in \mathcal{T}_i(x_i)$, one can check $x_i^k \uparrow x_i$ is not a good choice when $x_i < E$.\\

An alternative approach to payoff discontinuities induced by ties consists in allowing \emph{endogenous tie-breaking rules}, whereby the resolution of ties is not fixed ex-ante but is instead selected as part of the equilibrium outcome. Existence results for such games are provided by~\cite{SimonZame1990}, who study games with endogenous sharing rules and establish equilibrium existence under weak continuity requirements.

\vskip 0.5cm

\begin{theorem}{\textbf{(\cite{SimonZame1990})}}\\
Every compact metric game admits a mixed sharing rule solution.
\end{theorem}

\vskip 0.5cm

\subsection{Dasgupta and Maskin approach}

\hskip 0.5cm In the sequel, we establish an adapted version of Theorem 5 in~\cite{Dasgupta1986}. In order to do so, we start by adopting the same formalism.\\

We define hereafter, $\forall i$,
\begin{align*}
\mathbb{T}_i&=\left\{(x_1,\hdots,x_N)\in [A,B]^N \mid \exists j\neq i , \quad x_j=f^0_{ij}(x_i)=x_i  \right\} \\
\mathbb{F} _i&=\left\{(x_1,\hdots,x_N)\in [A,B]^N \mid \exists j\neq i , \quad x_j=f^1_{ij}(\mathbf{x}_{-j})= (2N-1)x_i - \sum_{k\neq i,j} x_k -N\,E  \right\} \\
\mathbb{P} _i&=\left\{(x_1,\hdots,x_N)\in [A,B]^N \mid \exists j\neq i , \quad x_j=f^2_{ij}(\mathbf{x}_{-j})= \frac{\sum_{k\neq j} x_k + N E}{2N-1} \right\} 
\end{align*}

The three sets have $N$-Lebesgue measure zero and satisfy $\mathcal{T}_i \subset \mathbb{T}_i$, $\mathcal{F}_i \subset \mathbb{F}_i$ and $\mathcal{P}_i \subset \mathbb{P}_i$. Define, for all $z_i \in [A,B]$,
\begin{align*}
\mathcal{T}_{-i}(z_i) &= \{ \mathbf{x}_{-i} \in [A,B]^{N-1} \mid (z_i, x_{-i}) \in \mathcal{T}_i \}, \\
\mathcal{F}_{-i}(z_i) &= \{ \mathbf{x}_{-i} \in [A,B]^{N-1} \mid (z_i, x_{-i}) \in \mathcal{F}_i \}, \\
\mathcal{P}_{-i}(z_i) &= \{ \mathbf{x}_{-i} \in [A,B]^{N-1} \mid (z_i, x_{-i}) \in \mathcal{P}_i \}, \\
\mathbb{T}_{-i}(z_i) &= \{ \mathbf{x}_{-i} \in [A,B]^{N-1} \mid (z_i, x_{-i}) \in \mathbb{T}_i \}, \\
\mathbb{F}_{-i}(z_i) &= \{ \mathbf{x}_{-i} \in [A,B]^{N-1} \mid (z_i, x_{-i}) \in \mathbb{F}_i \},\\
\mathbb{P}_{-i}(z_i) &= \{ \mathbf{x}_{-i} \in [A,B]^{N-1} \mid (z_i, x_{-i}) \in \mathbb{P}_i \}.
\end{align*}

We denote by $\mathcal{T}_i(i)$ (resp. $\mathcal{F}_i(i)$, $\mathcal{P}_i(i)$, $\mathbb{T}_i(i)$, $\mathbb{F}_i(i)$, $\mathbb{P}_i(i)$) the projection of $\mathcal{T}_i$ (resp. $\mathcal{F}_i$, $\mathcal{P}_i$, $\mathbb{T}_i$, $\mathbb{F}_i$, $\mathbb{P}_i$) onto $[A,B]$.

\vskip 0.5cm

\begin{remark}{\ }\\
For $N=2$, $\mathcal{F}_i$ becomes simple and adhere completely to the definition in~\cite{Dasgupta1986} of discontinuous sets.
\end{remark}

\vskip 0.5cm

\begin{lemma}{\ \label{Lemma1}}\\
Let $\mu_i$ be a Borel probability measure defined on $[A,B]$. If for all $i$, $\mu_i$ is atomless on $\mathcal{T}_i$ (resp. on $\mathcal{F}_i$, on $\mathcal{P}_i$), then $\boldsymbol{\mu}(\mathcal{T}_i) = 0$ (resp. $\boldsymbol{\mu}(\mathcal{F}_i) = 0$, $\boldsymbol{\mu}(\mathcal{P}_i) = 0$).\\

\noindent Furthermore, if for $z_i \in [A,B]$, $\mu_j(\{f^0_{ij}(z_i)\}) = 0$ for all $j \neq i$ (resp. $\mu_j(\{f^1_{ij}(z_i)\}) = 0$, $\mu_j(\{f^2_{ij}(z_i)\}) = 0$ for some $j \neq i$), then 
\begin{align*}
\boldsymbol{\mu}_{-i}(\mathbb{T}_i(z_i)) = 0, \quad (\text{resp.} \quad \boldsymbol{\mu}_{-i}(\mathbb{F}_i(z_i)) = 0, \quad \boldsymbol{\mu}_{-i}(\mathbb{P}_i(z_i)) = 0). 
\end{align*}

\end{lemma}

\vskip 0.5cm

\begin{proof}{\ }\\
In the sequel, we deal with $\mathcal{F}_i$ and $\mathbb{F}_i$, since the ties situation $\mathcal{T}_i$ and $\mathbb{T}_i$ is already proved in~\cite{Dasgupta1986}, the prove for the sets $\mathcal{P}_i$ and $\mathbb{P}_i$ is similar. By definition of $\mathcal{F}_i$, one can fix $j\neq i$ to write
\begin{align}
\boldsymbol{\mu}(\mathcal{F}_i) &= (\mu_1 \times \hdots \times \mu_N) (\mathcal{F}_i) \\
& \leq (\mu_1 \times \hdots \times \mu_N) \Big( (x_1, \dots, x_N) \in [A,B]^N \ \big| \  x_j = f^1_{ij}(\mathbf{x}_{-j}) \Big) \cap \mathcal{F}_i . \label{eq:7}
\end{align}

For each $j \neq i$, each positive integer $n$, and $r \in \{0, \dots, n-1\}$, define $\displaystyle a_r=A + \frac{r(B-A)}{n}$ then
\[
B^n_i(r) := \mathcal{F}_i(i) \cap \left\{ x_i \in [A,B] \ \bigg| \ a_r \le x_i \le a_{r+1} \right\},
\]
and
\[
B^n_j(\mathbf{r}_{-j}) := \mathcal{F}_j(j) \cap \left\{ x_j \in [A,B] \ \bigg| \ f^1_{ij}\Big( \mathbf{a}_{\mathbf{r}_{-j}} \Big) \le x_j \le f^1_{ij}\Big( \mathbf{a}_{\mathbf{r}_{-j}+1} \Big) \right\}.
\]

\noindent with $\mathbf{a}_{\mathbf{r}_{-j}}=(a_{r_1},\hdots,a_{r_{j-1}},a_{r_{j+1}},\hdots,a_{r_N})$. Because $\mu_i$ is atomless on $\mathcal{F}_i$ and due to the continuity of the function $f^1_{ij}$, for every $\varepsilon > 0$, there exists $n$ sufficiently large such that for all $k \neq j$, and $r_k \in \{0, \dots, n-1\}$,
\begin{align}
\mu_k(B^n_k(r_k)) < \varepsilon \quad \text{and} \quad \mu_j(B^n_j(\mathbf{r}_{-j})) < \varepsilon \quad \text{and} \quad \sum_{0\le \mathbf{r}_{-j} \le n-1} \mu_j(B^n_j(\mathbf{r}_{-j})) \leq \mu_j(\mathcal{F}_j(j)) \leq 1. \label{eq:8}
\end{align}

From \eqref{eq:7}, it follows that
\begin{align}
\label{eq:9}
\boldsymbol{\mu}(\mathcal{F}_i) 
&\le  \left\{ \sum_{0\le \mathbf{r}_{-j} \le n-1} \prod_{r_k \in \mathbf{r}_{-j}} \mu_k(B^n_k(r_k))\, \mu_j(B^n_j(\mathbf{r}_{-j})) \right\} < \varepsilon
\end{align}

Since $\varepsilon > 0$ is arbitrary, we conclude that
\begin{align*}
\boldsymbol{\mu}(\mathcal{F}_i) = 0.
\end{align*}

Next, suppose that for $z_i \in [A,B]$, $\mu_j(\{ f^1_{ij}(z_i,\mathbf{x}_{-{i,j}}) \}) = 0$ for some $j \neq i$. Then one can write
\begin{align*}
\mu_i(\mathbb{F}_{-i}(z_i)) 
&= \boldsymbol{\mu}_{-i} \Big( \{ \mathbf{x}_{-i} \in [A,B]^{N-1} \mid x_j=f^1_{ij}(z_i,\mathbf{x}_{-{i,j}}) \} \Big) \\
&= \mu_j \big( \{ f^1_{ij}(z_i,\mathbf{x}_{-{i,j}}) \} \big) \prod_{s\neq i,j} \mu_s \big( [A,B] \big) = 0,
\end{align*}
as desired.
\end{proof}

\vskip 0.5cm

Lemma 2 of~\cite{Dasgupta1986} holds without big differences.

\vskip 0.5cm

\begin{lemma}{\textbf{\cite{Dasgupta1986}\label{Lemma2}}\\}
Let $g_i : [A,B]^N \to \R$, $i = 1, \ldots, N$, be continuous except on a the closed subset $\mathcal{D}_i$ of $S$. Suppose $\forall i, \, g_i$ is bounded. Let $[A,B]_n$ be a finite subset of $[A,B]$ with the property that
\begin{align*}
\sup_{x_i \in [A,B]} \inf_{x_i^n \in [A,B]_n} |x_i - x_i^n| = \frac{1}{n}.
\end{align*}
Let $\boldsymbol{\mu}^n=(\mu^n, \ldots, \mu^n)$ be an equilibrium vector of mixed strategies for the symmetric finite game \mbox{$[([A,B]_n, g_i); \, i = 1, \ldots, N]$}, and let $\boldsymbol{\mu}^{\star} = \lim_{n \to \infty} \boldsymbol{\mu}^n$. If, for some $i$, there exists $y_i \in [A,B]$ such that
\begin{align*}
\int_{[A,B]^{N-1}} g_i(y_i, x_{-i}) \, d\boldsymbol{\mu}^{\star}_{-i} 
> \lim_{n \to \infty} \int_{[A,B]^{N-1}} g_i(x) \, d\boldsymbol{\mu}^n + \varepsilon,
\quad \text{for some } \varepsilon > 0,
\end{align*}
then 
\begin{align*}
\boldsymbol{\mu}^{\star}_{-i} \big(\mathcal{D}_{-i}(y_i) \big) > 0,
\end{align*}
\end{lemma}

\vskip 0.5cm

We now prove a modified version of Lemma 3 of~\cite{Dasgupta1986}.

\vskip 0.5cm

\begin{lemma}{\textbf{\cite{Dasgupta1986}\label{Lemma3}}\\}
Let $g_i : [A,B]^N \to \R$, $(i = 1, \ldots, N)$, be continuous except on the subset $\mathcal{D}_i$ of $[A,B]^N$. Suppose $\forall i$, $g_i$ is bounded. Let $[A,B]_n$ be a finite subset of $[A,B]$ with the property that
\begin{align*}
\sup_{x_i \in [A,B]} \inf_{x_i^n \in [A,B]_n} |x_i - x_i^n| < \frac{1}{n}.
\end{align*}
Let $\boldsymbol{\mu}^n=(\mu_1^n, \ldots, \mu_N^n)$ be an equilibrium vector of mixed strategies for the finite game $\{([A,B]_n, g_i);\, i = 1, \ldots, N\}$, and let 
$\boldsymbol{\mu}^{\star}=\lim_{n\to \infty} \boldsymbol{\mu}^n$. Then, for all but countably many $z_i \in [A,B]$,
\begin{align*}
\int_{[A,B]^{N-1}} g_i(z_i, \mathbf{x}_{-i}) \, d\boldsymbol{\mu}_{-i}^{\star} \leq 
\lim_{n \to \infty} \int_{[A,B]^N} g_i(\mathbf{x}) \, d\boldsymbol{\mu}^n,
\quad \forall i.
\end{align*}
\end{lemma}

\vskip 0.5cm

\begin{proof}{\ }\\
We prove the theorem here for $\mathcal{D}_i=\mathcal{F}_i$ (or similarly $\mathcal{D}_i=\mathcal{P}_i$), since the ties situation is already proved in \cite{Dasgupta1986}. Suppose that the hypotheses of the lemma hold, but that there exists $i$ such that
\begin{align}
\label{eqn:x}
\int_{[A,B]^{N-1}} g_i(z_i, \mathbf{x}_{-i}) \, d\boldsymbol{\mu}_{-i}^{\star} >
\lim_{n \to \infty} \int_{[A,B]^{N}} g_i(\mathbf{x}) \, d\boldsymbol{\mu}^n,
\quad \forall i.
\end{align}
for uncountably many $z_i \in [A,B]$. By Lemma \ref{Lemma2}, we conclude that for each $z_i \in [A,B]$ which satisfies \ref{eqn:x},
\begin{align*}
\boldsymbol{\mu}^{\star}_{-i} \big(\mathcal{F}_{-i}(z_i)\big) > 0.
\end{align*}

But then, by lemma \ref{Lemma1}, we know that for each such $z_i$, for all $j$ $(j \neq i)$ 
\begin{align}
\label{eqn:x2}
\mu^{\star}_j\big(\{ f^1_{ij}(z_i,\mathbf{x}_{-i,j}) \} \big) > 0.
\end{align}

\noindent for uncountably many $z_i$'s. Remember that $x_j$ is the solution set of the equation 
\begin{align*}
x_j&=f^1_{ij}(z_i,\mathbf{x}_{-{i,j}})= (2N-1)z_i - \sum_{k\neq i,j} x_k -N\,E 
\end{align*}

Or equivalently
\begin{align*}
\sum_{k\neq i} x_k &=f^1_{ij}(z_i,\mathbf{x}_{-{i,j}})= (2N-1)z_i  -N\,E 
\end{align*}

The $\sum_{k\neq i} x_k$ should then take uncountable many values, in particular, there exists at least one $l \neq i$, for which many uncountable $x_l$ satisfying
\begin{align*}
x_l&=f^1_{il}(x_i,\mathbf{x}_{-{i,l}})= (2N-1)x_i - \sum_{k\neq i,l} x_k -N\,E 
\end{align*}

Choosing $j=l$ leads to $\mu^{\star}_l$ to have uncountably many atoms. This is an impossibility.
\end{proof}

\vskip 0.5cm

Lemma 4 of~\cite{Dasgupta1986} holds without major modifications.

\vskip 0.5cm

\begin{lemma}{\textbf{\cite{Dasgupta1986}\label{Lemma4}}\\}
Let $g_i : [A,B]^N \to \R$, $(i = 1, \ldots, N)$, be continuous except on the closed subset $\mathcal{D}_i$ of $[A,B]^N$. Suppose $\forall i$, $g_i(\cdot,\mathbf{x}_{-i})$ is bounded and weakly lower semicontinuous, and $\sum_{i=1}^N g_i$ is upper semicontinuous. Let $[A,B]_n$ be a finite subset of $[A,B]$ with the property that
\begin{align*}
\sup_{x_i \in [A,B]} \inf_{x_i^n \in [A,B]_n} |x_i - x_i^n| < \frac{1}{n}.
\end{align*}
Let $\boldsymbol{\mu}^n=(\mu_1^n, \ldots, \mu_N^n)$ be an equilibrium vector of mixed strategies for the finite game $\{([A,B]_n, g_i);\, i = 1, \ldots, N\}$, and let 
$\boldsymbol{\mu}^{\star}=\lim_{n\to \infty} \boldsymbol{\mu}^n$. Then,
\begin{align*}
\int_{[A,B]^N} g_i(\mathbf{x}) \, d\boldsymbol{\mu}^{\star} =
\lim_{n \to \infty} \int_{[A,B]^N} g_i(\mathbf{x}) \, d\boldsymbol{\mu}^n,
\quad \forall i.
\end{align*}
\end{lemma}

\vskip 0.5cm

Then we can deduce the main existence theorem exactly as in Theorem 5 of~\cite{Dasgupta1986},

\vskip 0.5cm

\begin{theorem}{\ }\\
The symmetric game $[([A,B]^N,g_i);\, i=1,\dots,N]$ possesses a mixed strategy equilibrium $\boldsymbol{\mu}^{\star}$ and could be approached by the sequence $\boldsymbol{\mu}^n$.
\end{theorem}

\vskip 0.5cm

One can easily check the following property of the game.

\vskip 0.5cm

\begin{definition}{\textbf{Strict weak lower semicontinuity on the diagonal~\cite{Dasgupta1986}}}\\
Let $g_i : [A,B]^N \to \R$ denote player $i$'s payoff function. We say that $g_i(x_i,\mathbf{x}_{-i})$ is \emph{strict weakly lower semicontinuous on the diagonal} if it is weakly lower semicontinuous, and for every $z \in \mathcal{D}_i$, there exists $\lambda \in [0,1]$ such that
\begin{align*}
g_i(z,\hdots,z)
\;<\;
\lambda \liminf_{x_i \to \overline{x}_i^-} g_i(z,\hdots,z)
\;+\;
(1-\lambda) \liminf_{x_i \to \overline{x}_i^+} g_i(z,\hdots,z).
\end{align*}
\end{definition}

\vskip 0.5cm

\begin{theorem}{\ }\\
The symmetric game $[([A,B]^N,g_i);\, i=1,\dots,N]$ possesses a symmetric mixed strategy equilibrium \mbox{$\boldsymbol{\mu}^{\star}=\left({\mu}^{\star}, \hdots, {\mu}^{\star} \right)$} with the property that $\forall i$, $\forall z \in \mathcal{D}_i$, ${\mu}^{\star}(\{z\})=0$.
\end{theorem}

\vskip 0.5cm

\section*{Declarations}

\subsection*{Conflict of interest}

The author declares that they have no conflict of interest.

\subsection*{Authors' contributions}

The author carried out the study conception and design. Material preparation, data collection and analysis were performed by the author. The manuscript was written by the author.

\subsection*{Funding}

No funding was received for conducting this study.

\subsection*{Availability of data and materials}

The data used comes from Moroccan Public Procurement Portal freely accessible at :\\
\mbox{\url{https://www.marchespublics.gov.ma}}.

\subsection*{Declaration of generative AI and AI-assisted technologies in the writing process}

During the preparation of this work the author used ChatGPT in order to improve the text. After using this tool/service, the author reviewed and edited the content as needed and takes full responsibility for the content of the publication.

\bibliographystyle{apalike}
\bibliography{BibliographieEco}

\end{document}